\documentclass[11pt]{article}

\usepackage{mathrsfs,amsbsy,amssymb,latexsym,amsfonts,amsmath}
\usepackage[dvipdfm]{graphicx,color}
\usepackage{wrapfig}
\usepackage{amsbsy}
\usepackage{alltt}

\usepackage{array}

\newcommand{\Slash}[1]{\ooalign{\hfil/\hfil\crcr$#1$}}

\def\tr{\mathop{\rm tr}}

\makeatletter
	
	\@addtoreset{equation}{section}
\makeatother

\begin{document}

\begin{flushright}
OIQP-15-03
\end{flushright}


\begin{center}
{\Large \bf A state description of pair production from Dirac sea\\in gravitational field\\--physical interpretation of Weyl anomaly--}
\end{center}




\centerline{Yoshinobu {\sc Habara}$^a$, Holger B. {\sc Nielsen}$^b$ and Masao {\sc Ninomiya}$^a$}
\medskip
\medskip
\centerline{\small $^a$Okayama Institute for Quantum Physics, 1-9-1 Kyoyama, Okayama 700-0015, Japan}
\centerline{\small $^b$Niels Bohr Institute, University of Copenhagen, 17 Blegdamsvej, DK2100, Denmark}
\medskip
\medskip

\noindent
{\small {\bf E-mails:} habara@yukawa.kyoto-u.ac.jp, hbech@nbi.dk, msninomiya@gmail.com}


\begin{abstract}
We present the method that the pair production of particles is described as a Fock space state vector, where the pair production stems from receiving energy from, as an example, gravitational background field. At the same time we show a sort of new mechanism that in the diffeomorphism-invariant Dirac sea the Weyl symmetry is broken due to the pair production and this provides us a new physical interpretation of the origin of Weyl anomaly. In the present paper at the first place, we consider two kinds of background space-time which are connected each other by the Weyl transformation. It is shown that the vacuum, i.e. Dirac sea, in the ``curved" space-time \underline{after} making Weyl transformation is nothing but the pair produced state provided we consider the state in terms of the Fock space basis in the ``flat" space-time \underline{before} making the Weyl transformation. When summing up with respect to the time the contribution of pair production of each time-slice we obtain the Ricci scalar $R$ and then it is related to the trace part $\left\langle T^{\mu}_{\> \mu}\right\rangle$ of the energy-momentum tensor to produce Weyl anomaly.
\end{abstract}





\section{Introduction}\label{sec0}

It is well known that the virtual pair production and destruction occur always when the external field acts to the vacuum. Further these virtual pair produced particles gain the energy from the external field and then these particles appear as real particles. In quantum electrodynamics (QED) or similar, in particular the case of electron-positron in the presence of the external electric field this is known as the Schwinger effect~\cite{rf:schwinger} and has been well analyzed theoretically.

So far the analysis have been done by calculating the vacuum energy density $E$ and then the imaginary part ${\rm Im} E$ is considered as the particle-antiparticle pair production rate defined as 
\begin{align*}
	\Gamma =2{\rm Im} E.
\end{align*}
It is physically interpreted that the pair production is due to the tunnel effect, because the energy has the imaginary part.

\begin{figure}[htb]
\begin{center}
\includegraphics[width=70mm,angle=270]{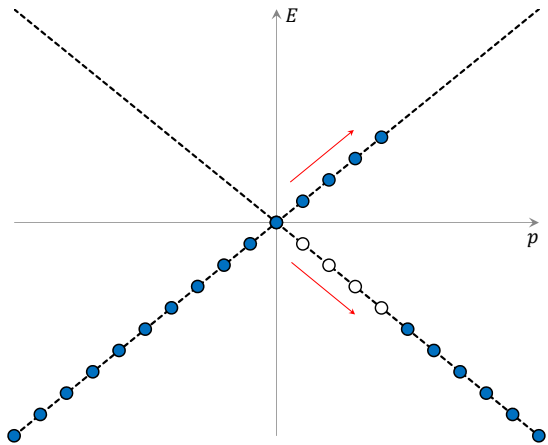}
\end{center}
\caption{The pair production due to being pumped up or down from the Dirac sea in \cite{rf:chiral}.}
\label{pic:pair.prod-0}
\end{figure}

A direct physical clarification of the pair production in the background field was made by two of us of the present authors (H.B.N. and M.N.)~\cite{rf:chiral}. When the electric field is applied to the negative energy electrons in the Dirac sea, the right-mover particles and left-mover holes gain the energy and are pumped up or down, the pair production occurs. This mechanism is depicted in Figure~\ref{pic:pair.prod-0} where the blue circle denotes the electron and the white circle is the positron.

The purpose of the present article is twofold. The first purpose is to reveal in a quantum field theoretical method the dynamics of the pair production. We show in more detail and precisely that the pair production state is a superposition of all possible particle and antiparticle state. In fact we will show that the pair production state is indeed a superposition of one pair, two pairs, three pairs, ... of particle and antiparticle with various energies. 

The second theme of the present paper is a kind of new physical derivation of the Weyl anomaly. In 1974, D. M. Capper and M. J. Duff discovered the Weyl anomaly~\cite{rf:capper.duff} that the trace of the energy-momentum tensor for a massless Weyl fermion in a gravitational background is not as classically expected to be $0$, but rather 
\begin{align}
	\langle T_{\mu}^{\> \mu}\rangle \propto \sqrt{g}R.
	\label{eq:0-weyl.anomaly}
\end{align}
Here $\sqrt{g}$ is the square root of the determinant of the metric tensor and $R$ is the Ricci scalar. Later many recalculations of this anomaly has been made (see the review, for example \cite{rf:duff:1993wm}). The present article is an attempt to obtain a physical understanding of how this anomaly comes about, i.e. to reveal the mechanism that the pair production from the Dirac sea leads the Weyl anomaly. As such it is also inspired by the interpretation of the chiral anomaly as pumping up or down states from or to the Dirac sea \cite{rf:chiral} of two of us (M.N and H.B.N). An another attempt to understand the $1+1$ dimensional Weyl anomaly by our thinking on the Dirac sea we presented in \cite{rf:copenhagen.version}.

As setting up we consider $1+1$ dimensional Euclidean cylinder type space-time with the conformal flat metric $g^{\prime}_{\mu \nu}=e^{2\Omega^{\prime}(\vec{x})}\eta_{\mu \nu}$. In this space-time we investigate the vacuum-to-vacuum transition amplitude of the massless Dirac fermion which has Weyl symmetric action. We then make a basis transformation of (or, in other words, pull back) the amplitude into the Fock space in the space-time with another metric $g_{\mu \nu}=e^{2\Omega (\vec{x})}\eta_{\mu \nu}$. Then we can show that the Dirac sea in the space-time specified by the primed metric $g^{\prime}_{\mu \nu}$ ranging from $-\infty$-past to $+\infty$-future is expressed on the basis in the space-time with $g_{\mu \nu}$ as a pair produced state on each time-slice. By summing up these pair production contribution from $-\infty$-past to $+\infty$-future, the difference between the vacuum (or Casimir) energy in the space-time of the \underline{primed} metric $g^{\prime}_{\mu \nu}$ and that of the \underline{unprimed} metric $g_{\mu \nu}$ gives rise to the Ricci scalar. In this way we finally obtain the Weyl anomaly.

The organization of this paper is as follows. In the following section \ref{sec1} we summarize the construction of the Weyl particles and Dirac sea in the gravitational background. In the section \ref{sec2} we then formulate the state description of the pair production and evaluate the Weyl anomaly using the Dirac sea formalism. In section \ref{sec3} we conclude and give an outlook as to how possibly to extend the use of our formalism. We have in addition three appendices.

\section{Complex fermion and the Dirac sea}\label{sec1}

In order to describe pair production state as a quantum state in a Fock space, we consider a complex fermion 
\begin{align}
	\Psi =\left( \begin{array}{c} \psi \\ \bar{\psi} \end{array} \right)
	\label{eq:1-fermion}
\end{align}
on the 2-dimensional Euclidean cylinder with 
\begin{align*}
	\vec{x}=(x^0,x^1)=(t,x), \quad \text{with }
	-\infty<t<+\infty \quad \text{and} \quad 0\leq x<2\pi .
\end{align*}
We adopt the notations that the Roman indices $i,j,\cdots$ and Greek indices $\mu ,\nu ,\cdots$ represent flat and curved space-time respectively. The diffeomorphism-invariant action for $\Psi$ reads 
\begin{align}
	S[\Psi,g] 
	& =\frac{1}{2\pi}\int d^2\vec{x}\sqrt{g}\> \Psi^{\dagger}(\vec{x})
	\gamma^0\gamma^ie_i^{\> \mu}(\vec{x})\nabla_{\mu}\Psi (\vec{x}) 
	\nonumber \\
	& =\frac{1}{4\pi}\int d^2\vec{x}\sqrt{g}\> 
	\left\{\Psi^{\dagger}(\vec{x})\gamma^0\gamma^ie_i^{\> \mu}(\vec{x})
	\nabla_{\mu}\Psi (\vec{x})-\nabla_{\mu}\Psi^{\dagger}(\vec{x})\cdot 
	\gamma^0\gamma^ie_i^{\> \mu}(\vec{x})\Psi (\vec{x}) \right\} 
	\nonumber \\
	& \equiv \int d^2\vec{x}\> L[g,\Psi]. 
	\label{eq:1-diffeo.action}
\end{align}
By using the degrees of freedom of diffeomorphism, we may take the metric tensor as the following conformal flat form 
\begin{align}
	g_{\mu \nu}=e^{2\Omega (\vec{x})}\eta_{\mu \nu}, 
	\label{eq:1-conformal.metric}
\end{align}
and thus the zweibeins read 
\begin{align}
	e_{00}=e_{11}=e^{\Omega (\vec{x})},\> e_{01}=e_{10}=0. 
	\label{eq:1-zweibein}
\end{align}
In this article we always use the conformal metric (\ref{eq:1-conformal.metric}) as the background space-time. We assume that the space-time is flat both for infinite past and future, i.e. 
\begin{align*}
	\lim_{t\to \pm \infty}g_{\mu \nu}(\vec{x})=\eta_{\mu \nu}
	=\left( \begin{array}{cc}
	1 & 0 \\ 0 & 1 
	\end{array} \right) 
\end{align*}
so that 
\begin{align*}
	\lim_{t\to \pm \infty}\Omega (\vec{x})=0. 
\end{align*}
Under the background space-time (\ref{eq:1-conformal.metric}) the action (\ref{eq:1-diffeo.action}) becomes 
\begin{align}
	S[\Psi,g]=\frac{1}{4\pi}\int d^2\vec{x}\> \left\{
	\Big(e^{\frac{1}{2}\Omega (\vec{x})}\Psi^{\dagger}\Big)\gamma^0\gamma^i
	\partial_i\Big(e^{\frac{1}{2}\Omega (\vec{x})}\Psi \Big)
	-\partial_i\Big(e^{\frac{1}{2}\Omega (\vec{x})}\Psi^{\dagger}\Big)
	\cdot \gamma^0\gamma^i
	\Big(e^{\frac{1}{2}\Omega (\vec{x})}\Psi \Big)\right\}. 
	\label{eq:1-action}
\end{align}
Here $e(\vec{x})\equiv \det e_{i\mu}(\vec{x})$ and thus $e^{\frac{1}{2}\Omega (\vec{x})}=\sqrt[8]{g}=\sqrt[4]{e}$. The action (\ref{eq:1-action}) apparently possesses the following Weyl invariance, 
\begin{align}
	\left\{ \begin{array}{l}
	\Omega (\vec{x})\to 
	\Omega^{\prime}(\vec{x})=\Omega (\vec{x})+\omega (\vec{x}) \\
	\Psi (\vec{x})\to \Psi^{\prime}(\vec{x})
	=e^{-\frac{1}{2}\omega (\vec{x})}\Psi (\vec{x})
	\end{array} \right.. 
	\label{eq:1-weyl.invariance}
\end{align}
Here we assumed the variation of the metric denoted by $\omega (\vec{x})$ also satisfies 
\begin{align*}
	\lim_{t\to \pm \infty}\omega (\vec{x})=0. 
\end{align*}

We investigate the Weyl transformation of the Hamiltonian density. The Hamiltonian density is given from the action (\ref{eq:1-diffeo.action}) as 
\begin{align}
	H(\vec{x}) & =T_{00}(\vec{x}) \nonumber \\
	& =\partial_0\Psi \frac{\partial L}{\partial (\partial_0\Psi)}
	+\partial_0
	\Psi^{\dagger}\frac{\partial L}{\partial (\partial_0\Psi^{\dagger})}-L 
	\nonumber \\
	& =\frac{1}{4\pi}e^{\Omega}\Psi^{\dagger}\partial_0\Psi 
	-\frac{1}{4\pi}e^{\Omega}\partial_0\Psi^{\dagger}\cdot \Psi-L 
	\nonumber \\
	& \equiv H[\Psi,g](\vec{x}). 
	\label{eq:1-hamiltonian.density}
\end{align}
The form (\ref{eq:1-hamiltonian.density}) transforms under the Weyl transformation (\ref{eq:1-weyl.invariance}) as 
\begin{align}
	H^{\prime}
	& =\frac{1}{4\pi}e^{\Omega +\omega}
	\Big(e^{-\frac{1}{2}\omega}\Psi^{\dagger}\Big)\partial_0
	\Big(e^{-\frac{1}{2}\omega}\Psi \Big)
	-\frac{1}{4\pi}e^{\Omega +\omega}
	\partial_0\Big(e^{-\frac{1}{2}\omega}\Psi^{\dagger}\Big)\cdot 
	\Big(e^{-\frac{1}{2}\omega}\Psi \Big)-L^{\prime} \nonumber \\
	& =\frac{1}{4\pi}e^{\Omega}\Psi^{\dagger}\partial_0\Psi 
	+\frac{1}{4\pi}e^{\Omega}\Psi^{\dagger}\Psi 
	\left(-\frac{1}{2}\partial_0\omega \right)
	-\frac{1}{4\pi}e^{\Omega}\partial_0\Psi^{\dagger}\cdot \Psi 
	-\frac{1}{4\pi}e^{\Omega}\Psi^{\dagger}\Psi 
	\left(-\frac{1}{2}\partial_0\omega \right)-L \nonumber \\
	& =H. 
	\label{eq:1-hamiltonian.density.weyl}
\end{align}
This shows that $H$ is Weyl invariant. 

We then quantize the system by finding the solutions for $\psi$ and $\bar{\psi}$. The equations of motion from (\ref{eq:1-action}) read 
\begin{align}
	\left( \begin{array}{cc}
	\partial_0+i\partial_1 & 0 \\
	0 & \partial_0-i\partial_1
	\end{array} \right)
	\left( \begin{array}{c}
	e^{\frac{1}{2}\Omega}\psi \\
	e^{\frac{1}{2}\Omega}\bar{\psi}
	\end{array} \right)=0, \quad 
	\left( \begin{array}{cc}
	\partial_0+i\partial_1 & 0 \\
	0 & \partial_0-i\partial_1
	\end{array} \right)
	\left( \begin{array}{c}
	e^{\frac{1}{2}\Omega}\psi^{\dagger} \\
	e^{\frac{1}{2}\Omega}\bar{\psi}^{\dagger}
	\end{array} \right)=0 
	\label{eq:1-eom}
\end{align}
and when imposing the periodic boundary condition 
\begin{align*}
	& \psi (t,x+2\pi )=\psi (t,x), \\
	& \bar{\psi}(t,x+2\pi )=\bar{\psi}(t,x), 
\end{align*}
the general solution is given by 
\begin{align}
	& \psi (\vec{x})
	=\frac{1}{\sqrt{2\pi}}\sum_{n=-\infty}^{+\infty}b_n^{[\Psi,\Omega]}
	\frac{e^{-\frac{1}{2}\Omega (\vec{x})}}{e^{n(t+ix)}}, \quad 
	\psi^{\dagger}(\vec{x})
	=\frac{1}{\sqrt{2\pi}}\sum_{n=-\infty}^{+\infty}
	b_n^{[\Psi,\Omega]\dagger}
	e^{-\frac{1}{2}\Omega (\vec{x})}e^{n(t+ix)}, \nonumber \\
	& \bar{\psi}(\vec{x})
	=\frac{1}{\sqrt{2\pi}}\sum_{n=-\infty}^{+\infty}
	\bar{b}_n^{[\Psi,\Omega]}
	\frac{e^{-\frac{1}{2}\Omega (\vec{x})}}{e^{n(t-ix)}}, \quad 
	\bar{\psi}^{\dagger}(\vec{x})
	=\frac{1}{\sqrt{2\pi}}\sum_{n=-\infty}^{+\infty}
	\bar{b}_n^{[\Psi,\Omega]\dagger}
	e^{-\frac{1}{2}\Omega (\vec{x})}e^{n(t-ix)}. 
	\label{eq:1-solution}
\end{align}
The suffix $[\Psi,\Omega]$ on the expanding coefficients in (\ref{eq:1-solution}) which represents the original field $\Psi$ and the background metric $g_{\mu \nu}=e^{2\Omega}\eta_{\mu \nu}$ is added for later use. Thus the eigenfunctions, i.e. the  basis of the Hilbert space read 
\begin{align}
	& f_n^{[\Omega]}(\vec{x})=\frac{1}{\sqrt{2\pi}}
	\frac{e^{-\frac{1}{2}\Omega (\vec{x})}}{e^{n(t+ix)}}, \quad 
	f_n^{[\Omega]\dagger}(\vec{x})
	=\frac{1}{\sqrt{2\pi}}e^{-\frac{1}{2}\Omega (\vec{x})}e^{n(t+ix)}, 
	\nonumber \\
	& \bar{f}_n^{[\Omega]}(\vec{x})=\frac{1}{\sqrt{2\pi}}
	\frac{e^{-\frac{1}{2}\Omega (\vec{x})}}{e^{n(t-ix)}}, \quad 
	\bar{f}_n^{[\Omega]\dagger}(\vec{x})
	=\frac{1}{\sqrt{2\pi}}e^{-\frac{1}{2}\Omega (\vec{x})}e^{n(t-ix)}.
	\label{eq:1-basis}
\end{align}
Here the hermitian conjugation $\dagger$ is understood as the complex conjugation in Hilbert space, after making Wick rotation $\tau =-it$, 
\begin{align*}
	\{e^{in(\tau \pm x)}\}^{\dagger}=e^{-in(\tau \pm x)}. 
\end{align*}
The inner product in the Hilbert space is defined on the time-slice $t$ as 
\begin{align}
	& \langle n|m\rangle 
	=\int dx\> \langle n|x\rangle_{[\Omega]}
	{}_{[\Omega]}\langle x|m\rangle 
	=\oint dx\> \sqrt{g}
	f_n^{[\Omega]\dagger}(\vec{x})e_0^{\> 0}f_m^{[\Omega]}(\vec{x})
	=\delta_{n,m}, \nonumber \\
	& \langle \bar{n}|\bar{m}\rangle 
	=\int dx\> \langle \bar{n}|x\rangle_{[\Omega]}
	{}_{[\Omega]}\langle x|\bar{m}\rangle 
	=\oint dx\> \sqrt{g}
	\bar{f}_n^{[\Omega]\dagger}(\vec{x})e_0^{\> 0}
	\bar{f}_m^{[\Omega]}(\vec{x})=\delta_{n,m}. 
	\label{eq:1-inner.product}
\end{align}
Here the zweibein component 
\begin{align*}
	e_0^{\> 0}=e^{-\Omega (\vec{x})}
\end{align*}
comes from the diagonal component of the matrix 
\begin{align*}
	\gamma^0\gamma^ie_i^{\> 0}=\begin{pmatrix}1&0\\0&1\end{pmatrix}e_0^{\> 0}
\end{align*}
which makes the inner product covariant. 

We make the canonical second quantization 
\begin{align}
	& \left\{\psi (t,x),\> \psi^{\dagger}(t,x^{\prime})\right\}
	=\frac{1}{\sqrt{e(\vec{x})}}\delta (x-x^{\prime}), \nonumber \\
	& \left\{\bar{\psi}(t,x),\> \bar{\psi}^{\dagger}(t,x^{\prime})
	\right\}=\frac{1}{\sqrt{e(\vec{x})}}\delta (x-x^{\prime})
	\label{eq:1-canonical.quantization}
\end{align}
and then we have 
\begin{align}
	\left\{b_n^{[\Psi,\Omega]},b_m^{[\Psi,\Omega]\dagger}\right\}
	=\left\{\bar{b}_n^{[\Psi,\Omega]},\bar{b}_m^{[\Psi,\Omega]\dagger}
	\right\}
	=\delta_{n,m}, \quad \text{others}=0.
	\label{eq:1-commutator}
\end{align}
We should notice that the commutation relation (\ref{eq:1-commutator}) does not depend on the conformal factor function $\Omega (\vec{x})$ of the metric (\ref{eq:1-conformal.metric}). In other word the algebras of the creation and annihilation of the particle are all isomorphic, independent of $\Omega (\vec{x})$.

\begin{figure}[htb]
\begin{center}
\includegraphics[width=80mm,angle=270]{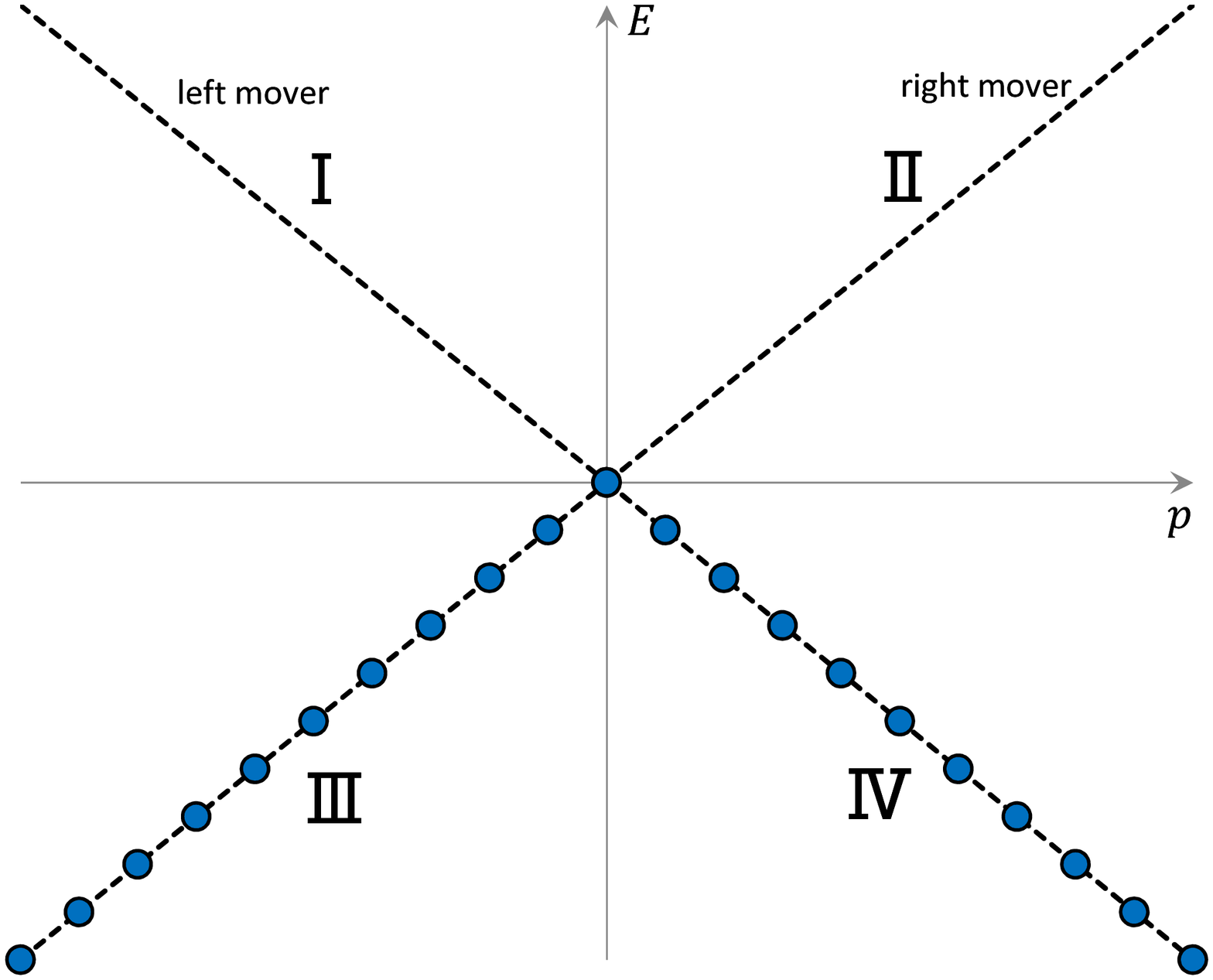}
\end{center}
\caption[Dirac sea]{The particle distribution of the Dirac sea}
\label{pic:dirac}
\end{figure}

In the present article we consider the Euclidean space-time, and in particular in the case of the flat space-time $\Omega (\vec{x})=0$, the energy operator is given by $-\frac{d}{dt}$, and the momentum one is $-i\frac{d}{dx}$. Thus e.g. the eigenstate $e^{-n(t\pm ix)}$ has the energy and momentum eigenvalues $E=n$ and $p=\mp n$ respectively. Therefore left mover $\psi (\vec{x})$ has the dispersion relation $E=-p$ while the right mover $\bar{\psi}(\vec{x})$ has $E=p$. From the above consideration, the vacuum state, i.e. Dirac sea is generically defined as 
\begin{align}
	|\text{sea}\rangle_{[\Psi,\Omega]}=
	\prod_{n\geq 0}b_n^{[\Psi,\Omega]\dagger}|0\rangle \otimes 
	\prod_{m<0}\bar{b}_m^{[\Psi,\Omega]\dagger}|\bar{0}\rangle 
	\label{eq:1-dirac.sea}
\end{align}
such that 
\begin{align*}
	b_n^{[\Psi,\Omega]}|0\rangle =\bar{b}_n^{[\Psi,\Omega]}|\bar{0}\rangle 
	=0 \quad \text{for all $n$. }
\end{align*}
It should be noticed that the index $n$ in $b_n^{[\Psi,\Omega]}$ and $\bar{b}_n^{[\Psi,\Omega]}$ is the momentum. Needless to say the statement ``the Dirac sea is the state filled by the negative energy particles" is meaningful only when $\Omega (\vec{x})=0$, i.e. the background space-time is flat. In this case the configuration of the particles in the Dirac sea is depicted by Figure~\ref{pic:dirac}. 
The operators $b_n^{[\Psi,0]}, b_n^{[\Psi,0]\dagger}, \bar{b}_n^{[\Psi,0]}$ and $\bar{b}_n^{[\Psi,0]\dagger}$ are the creation- and  annihilation-operators of the particles as is shown in Table \ref{tab:spectrum}. 
\begin{table}
\begin{center}
\extrarowheight=-2pt
\def\arraystretch{2}
\begin{tabular}{|@{~~}l@{~~}|@{~~}l@{~~}|}\hline
	spectrum I 
	& $b_{n<0}^{[\Psi,0]\dagger}$ create a positive-energy particle, \\
	& $b_{n<0}^{[\Psi,0]}$ annihilate a positive-energy particle, \\ \hline
	spectrum II 
	& $\bar{b}_{n\geq 0}^{[\Psi,0]\dagger}$ 
	create a positive-energy particle, \\
	& $\bar{b}_{n\geq 0}^{[\Psi,0]}$ 
	annihilate a positive-energy particle, 
	\\ \hline
	spectrum III 
	& $\bar{b}_{n<0}^{[\Psi,0]\dagger}$ create a negative-energy particle 
	\\
	& \hspace{1cm} (annihilate a positive-energy hole), \\
	& $\bar{b}_{n<0}^{[\Psi,0]}$ annihilate a negative-energy particle \\
	& \hspace{1cm} (create a positive-energy hole). \\ \hline
	spectrum IV 
	& $b_{n\geq 0}^{[\Psi,0]\dagger}$ create a negative-energy particle \\
	& \hspace{1cm} (annihilate a positive-energy hole), \\
	& $b_{n\geq 0}^{[\Psi,0]}$ annihilate a negative-energy particle \\
	& \hspace{1cm} (create a positive-energy hole), \\ \hline
\end{tabular}
	\caption{The interpretation of the creation and annihilation operators}
	\label{tab:spectrum}
\end{center}
\end{table}

In order to reveal the transformation property of the Dirac sea under the Weyl transformation, we consider the functional representation of the algebra (\ref{eq:1-commutator})~\cite{rf:hole-th}. As is easily understood the commutation relation (\ref{eq:1-commutator}) is expressed in terms of the Grassmannian c-numbers $\theta_n^{[\Psi,\Omega]}$ and $\bar{\theta}_n^{[\Psi,\Omega]}$ as 
\begin{align}
	& \left\{ \begin{array}{ll}
	b_n^{[\Psi,\Omega]\dagger}=\theta_n^{[\Psi,\Omega]}, 
	& b_n^{[\Psi,\Omega]}
	=\frac{\partial}{\partial \theta_n^{[\Psi,\Omega]}} \\
	\bar{b}_n^{[\Psi,\Omega]\dagger}=\bar{\theta}_n^{[\Psi,\Omega]}, 
	& \bar{b}_n^{[\Psi,\Omega]}
	=\frac{\partial}{\partial \bar{\theta}_n^{[\Psi,\Omega]}}
	\end{array} \right.. 
	\label{eq:1-fermion.rep}
\end{align}
Then the states in (\ref{eq:1-dirac.sea}) are written as 
\begin{align}
	& \text{empty vacua: }\quad |0\rangle \simeq 1,\> 
	|\bar{0}\rangle \simeq 1, \nonumber \\
	& \text{Dirac sea: }\quad |\text{sea}\rangle_{[\Psi,\Omega]}\simeq 
	\prod_{n\geq 0,\> m<0}\theta_n^{[\Psi,\Omega]}
	\bar{\theta}_m^{[\Psi,\Omega]}.
	\label{eq:1-dirac.sea.rep}
\end{align}
In conclusion, the empty vacua $|0\rangle$ and $|\bar{0}\rangle$ do not include any wave function component, so that it is nothing but ``classical number" $1$. Therefore it does not transform under the Weyl transformation.

\section{Weyl anomaly from Dirac sea}\label{sec2}

\subsection{Identity for the Weyl transformation}\label{sec2.1}

In this section we investigate the vacuum-to-vacuum amplitude under the Weyl transformation (\ref{eq:1-conformal.metric}) in order to show that the Weyl anomaly actually stems from pair production from the Dirac sea. For later convenience we rescale the complex fermion field $\Psi$ by using the metric tensor in the following: 
\begin{align}
	\tilde{\Psi}=\left( \begin{array}{c}
	\tilde{\psi} \\ \tilde{\bar{\psi}} 	\end{array} \right) 
	\equiv \sqrt[4]{g}\left( \begin{array}{c}
	\psi \\ \bar{\psi} \end{array} \right). 
	\label{eq:2-rescale}
\end{align}
The rescaled field $\tilde{\Psi}$ is introduced so that the Dirac sea becomes invariant under diffeomorphism\footnote{Originally the rescaled field (\ref{eq:2-rescale}) was introduced by K. Fujikawa in order to make the path integral measure invariant under diffeomorphism.}. This invariance of the Dirac sea is shown in the Appendix \ref{app1}. Then the Weyl invariance (\ref{eq:1-weyl.invariance}) of the action (\ref{eq:1-action}) leads to 
\begin{align}
	S[\tilde{\Psi},g]=\frac{1}{2\pi}\int d^2\vec{x}\> 
	\left\{\Big(e^{-\frac{1}{2}\Omega (\vec{x})}\tilde{\Psi}^{\dagger}\Big)
	\gamma^0\gamma^i\partial_i\Big(e^{-\frac{1}{2}\Omega (\vec{x})}
	\tilde{\Psi}\Big)\right\}
	\label{eq:2-rescaled.action}
\end{align}
where 
\begin{align}
	& \Omega (\vec{x})\to 
	\Omega^{\prime}(\vec{x})=\Omega (\vec{x})+\omega (\vec{x}), 
	\nonumber \\
	& \tilde{\Psi}(\vec{x})\to 
	\tilde{\Psi}^{\prime}(\vec{x})=e^{\frac{1}{2}\omega (\vec{x})}
	\tilde{\Psi}(\vec{x}).
	\label{eq:2-weyl.trans}
\end{align}
Now, by denoting the time ordering operator by $T$ and using the expression 
\begin{align*}
	H^{\prime}=H[\tilde{\Psi}^{\prime},g^{\prime}] 
\end{align*}
(see (\ref{eq:1-hamiltonian.density}) and (\ref{eq:1-hamiltonian.density.weyl})), the following identity for the vacuum-to-vacuum amplitude holds, which will be shown soon below: 
\begin{align}
	& {}_{[\tilde{\Psi}^{\prime},\Omega^{\prime}]}\langle \text{sea}|
	Te^{-\int d^2\vec{x}\> H[\tilde{\Psi}^{\prime},g^{\prime}]}
	|\text{sea}\rangle_{[\tilde{\Psi}^{\prime},\Omega^{\prime}]} 
	\nonumber \\
	& ={}_{[\tilde{\Psi},\Omega]}\langle \text{sea}|
	Te^{-\int d^2\vec{x}\> H[\tilde{\Psi},g^{\prime}]}
	|\text{sea}\rangle_{[\tilde{\Psi},\Omega]} \nonumber \\
	& ={}_{[\tilde{\Psi},\Omega]}\langle \text{sea}|J_{\omega}\cdot 
	Te^{-\int d^2\vec{x}\> H[\tilde{\Psi},g]}
	|\text{sea}\rangle_{[\tilde{\Psi},\Omega]}. 
	\label{eq:2-identity}
\end{align}
This identity is reminiscent to the formula (7.13) in the Fujikawa and Suzuki's text \cite{rf:fujikawa} for the Fujikawa's method of anomalies. Here we should notice that the second suffixes $\Omega^{\prime}$ and $\Omega$ of $|\text{sea}\rangle_{[\tilde{\Psi}^{\prime},\Omega^{\prime}]}$ and $|\text{sea}\rangle_{[\tilde{\Psi},\Omega]}$ represent the metrics on which the Fock spaces are constructed. We denote the Fock spaces $\mathcal{H}_{\Omega^{\prime}}$ and $\mathcal{H}_{\Omega}$ that are built up from the vacua $|\text{sea}\rangle_{[\tilde{\Psi}^{\prime},\Omega^{\prime}]}$ and $|\text{sea}\rangle_{[\tilde{\Psi},\Omega]}$ respectively. Needless to say, the inner products and the hermitian conjugations are defined in $\mathcal{H}_{\Omega^{\prime}}$ and $\mathcal{H}_{\Omega}$ respectively. As will be described later in detail, the numerical factor $J_{\omega}$ in the third line will be found to be the Weyl anomaly itself which emerges, simply saying, from the difference between the Dirac seas $|\text{sea}\rangle_{[\tilde{\Psi}^{\prime},\Omega^{\prime}]}$ and $|\text{sea}\rangle_{[\tilde{\Psi},\Omega]}$. It is always possible to insert fields in these amplitudes. However it is not essential point so that we disregard on this issue. We may easily understand the reason why the identity (\ref{eq:2-identity}) holds by the following explanations: 
\begin{enumerate}
\item[(i)] \underline{The equality between the first and second line of (\ref{eq:2-identity})}

We may apply with recourse to the isomorphism of the algebras of the creation- and annihilation-operators in the sense that the commutation relations (\ref{eq:1-commutator}) do not depend on the metric. Thus we may replace the operators $b_n^{[\tilde{\Psi}^{\prime},\Omega^{\prime}]}$ by $b_n^{[\tilde{\Psi},\Omega]}$. In the path integral formalism this operation corresponds to the arbitrary change of the naming of the integration variables. In fact the states and the operators in each amplitudes can be written as 
\begin{align}
	& |\text{sea}\rangle_{[\tilde{\Psi}^{\prime},\Omega^{\prime}]}
	=\prod_{n\geq 0}b_n^{[\tilde{\Psi}^{\prime},\Omega^{\prime}]\dagger}
	|0\rangle \otimes 
	\prod_{m<0}\bar{b}_m^{[\tilde{\Psi}^{\prime},\Omega^{\prime}]\dagger}
	|\bar{0}\rangle , 
	\quad \text{for}\quad 
	\tilde{\psi}^{\prime}(\vec{x})=\frac{1}{\sqrt{2\pi}}\sum_{n}
	b_n^{[\tilde{\Psi}^{\prime},\Omega^{\prime}]}
	\frac{e^{\frac{1}{2}\Omega^{\prime}(\vec{x})}}{e^{n(t+ix)}}
	\quad \text{etc.}, \nonumber \\
	& |\text{sea}\rangle_{[\tilde{\Psi},\Omega]}
	=\prod_{n\geq 0}b_n^{[\tilde{\Psi},\Omega]\dagger}|0\rangle \otimes 
	\prod_{m<0}\bar{b}_m^{[\tilde{\Psi},\Omega]\dagger}|\bar{0}\rangle , 
	\quad \text{for}\quad 
	\tilde{\psi}(\vec{x})=\frac{1}{\sqrt{2\pi}}\sum_{n}
	b_n^{[\tilde{\Psi},\Omega]}
	\frac{e^{\frac{1}{2}\Omega (\vec{x})}}{e^{n(t+ix)}}
	\quad \text{etc.}. 
	\tag{\ref{eq:2-identity}a}
	\label{eq:2-identity_a}
\end{align}
As for time translation operators, we write them for the Hamiltonian itself (not for the Hamiltonian density), 
\begin{align}
	e^{-\int d^2\vec{x}\> H[\tilde{\Psi}^{\prime},g^{\prime}]}
	& =e^{-\int dt\> \mathcal{H}^{[\tilde{\Psi}^{\prime},
	\Omega^{\prime}]}}, \nonumber \\
	\text{where}\quad \mathcal{H}^{[\tilde{\Psi}^{\prime},\Omega^{\prime}]}
	& =\frac{1}{2\pi}\sum_{n=1}^{+\infty}n\left(
	b_{-n}^{[\tilde{\Psi}^{\prime},\Omega^{\prime}]\dagger}
	b_{-n}^{[\tilde{\Psi}^{\prime},\Omega^{\prime}]}
	+b_n^{[\tilde{\Psi}^{\prime},\Omega^{\prime}]}
	b_n^{[\tilde{\Psi}^{\prime},\Omega^{\prime}]\dagger}
	+\bar{b}_n^{[\tilde{\Psi}^{\prime},\Omega^{\prime}]\dagger}
	\bar{b}_n^{[\tilde{\Psi}^{\prime},\Omega^{\prime}]}
	+\bar{b}_{-n}^{[\tilde{\Psi}^{\prime},\Omega^{\prime}]}
	\bar{b}_{-n}^{[\tilde{\Psi}^{\prime},\Omega^{\prime}]\dagger}
	\right) \nonumber \\
	& \quad \> +E_0^{[\tilde{\Psi}^{\prime},\Omega^{\prime}]}, \nonumber \\
	e^{-\int d^2\vec{x}\> H[\tilde{\Psi},g^{\prime}]}
	& =e^{-\int dt\> \mathcal{H}^{[\tilde{\Psi},\Omega^{\prime}]}}, 
	\nonumber \\
	\text{where}\quad \mathcal{H}^{[\tilde{\Psi},\Omega^{\prime}]}
	& =\frac{1}{2\pi}\sum_{n=1}^{+\infty}n\left(
	b_{-n}^{[\tilde{\Psi},\Omega]\dagger}b_{-n}^{[\tilde{\Psi},\Omega]}
	+b_n^{[\tilde{\Psi},\Omega]}b_n^{[\tilde{\Psi},\Omega]\dagger}
	+\bar{b}_n^{[\tilde{\Psi},\Omega]\dagger}
	\bar{b}_n^{[\tilde{\Psi},\Omega]}
	+\bar{b}_{-n}^{[\tilde{\Psi},\Omega]}
	\bar{b}_{-n}^{[\tilde{\Psi},\Omega]\dagger}
	\right) \nonumber \\
	& \quad \> +E_0^{[\tilde{\Psi},\Omega^{\prime}]}. 
	\tag{\ref{eq:2-identity}b}
	\label{eq:2-identity_b}
\end{align}
In this way the Dirac seas and the normal ordered operators are given by the replacement that is the consequence of the isomorphism of the algebras (\ref{eq:1-commutator}). Note that the renormalized vacuum energies that are c-numbers fulfill the equality $E_0^{[\tilde{\Psi}^{\prime},\Omega^{\prime}]}=E_0^{[\tilde{\Psi},\Omega^{\prime}]}$ so that they are regularized both on the metric $g^{\prime}_{\mu \nu}(\vec{x})=e^{2\Omega^{\prime}(\vec{x})}\eta_{\mu \nu}$.

\item[(ii)] \underline{The equality between the first and third line of (\ref{eq:2-identity})}

It is proven by using the Weyl transformation (\ref{eq:2-weyl.trans}) in the following manner: As was shown in (\ref{eq:1-hamiltonian.density.weyl}), the Hamiltonian density is Weyl invariant 
\begin{align*}
	H[\tilde{\Psi}^{\prime},g^{\prime}]=H[\tilde{\Psi},g]. 
\end{align*}
We may write in the same manner (\ref{eq:2-identity_b}) as an explicit form, 
\begin{align}
	e^{-\int d^2\vec{x}\> H[\tilde{\Psi},g]}
	& =e^{-\int dt\> \mathcal{H}^{[\tilde{\Psi},\Omega]}}, \nonumber \\
	\text{where}\quad \mathcal{H}^{[\tilde{\Psi},\Omega]}
	& =\frac{1}{2\pi}\sum_{n=1}^{+\infty}n\left(
	b_{-n}^{[\tilde{\Psi},\Omega]\dagger}b_{-n}^{[\tilde{\Psi},\Omega]}
	+b_n^{[\tilde{\Psi},\Omega]}b_n^{[\tilde{\Psi},\Omega]\dagger}
	+\bar{b}_n^{[\tilde{\Psi},\Omega]\dagger}
	\bar{b}_n^{[\tilde{\Psi},\Omega]}
	+\bar{b}_{-n}^{[\tilde{\Psi},\Omega]}
	\bar{b}_{-n}^{[\tilde{\Psi},\Omega]\dagger}
	\right)+E_0^{[\tilde{\Psi},\Omega]}. 
	\tag{\ref{eq:2-identity}c}
	\label{eq:2-identity_c}
\end{align}
Here the metric is changed to $g_{\mu \nu}(\vec{x})=e^{2\Omega (\vec{x})}\eta_{\mu \nu}$ according to the Weyl transformation and the vacuum energy $E_0^{[\tilde{\Psi},\Omega]}$ is the one that is regularized on this metric. On the other hand, the initial and final states $|\text{sea}\rangle_{[\tilde{\Psi}^{\prime},\Omega^{\prime}]}$ as $t\to \pm \infty$ are given by the condition $\lim_{t\to \pm \infty}\omega (\vec{x})=0$, and thus are trivially equal to that on the flat space-time, 
\begin{align*}
	|\text{sea}\rangle_{[\tilde{\Psi}^{\prime},\Omega^{\prime}]}
	=|\text{sea}\rangle_{[\tilde{\Psi},\Omega]}. 
\end{align*}
Of course in $-\infty <t< +\infty$, Dirac sea $|\text{sea}\rangle_{[\tilde{\Psi}^{\prime},\Omega^{\prime}]}$ needs to be carefully treated. In fact during the time development from $t=-\infty$ to $t=+\infty$ the metric may change, so that the first line may produce some quantity expressed as $J_{\omega}$ (Note that this contribution corresponds to the Jacobian factor of the path integral measure in Fujikawa's method). We postpone to give an explicit form to the next subsection. There we will obtain $J_{\omega}$ by calculating a change of the Dirac sea $|\text{sea}\rangle_{[\tilde{\Psi}^{\prime},\Omega^{\prime}]}$ under the Weyl transformation. 

In the present subsection we make a preparation to the next subsection in the following. We divide the amplitude in the first line of (\ref{eq:2-identity}) in terms of an infinitesimal time interval $\Delta t$. We hereby insert the completeness relation $1=\sum_{n=0}^{\infty}|n\rangle_{[\tilde{\Psi}^{\prime},\Omega^{\prime}]}{}_{[\tilde{\Psi}^{\prime},\Omega^{\prime}]}\langle n|$ in the Fock space $\mathcal{H}_{\Omega^{\prime}}$, where the $n$-particle state $|n\rangle_{[\tilde{\Psi}^{\prime},\Omega^{\prime}]}$ denotes the basis of $\mathcal{H}_{\Omega^{\prime}}$ and the intermediate state on the each time-slice. In this case, the initial and final states are vacua $|0\rangle_{[\tilde{\Psi}^{\prime},\Omega^{\prime}]}=|\text{sea}\rangle_{[\tilde{\Psi}^{\prime},\Omega^{\prime}]}$ so that when we perform summation over all the $n$-particle states on the time-slice at $t$, the contributing intermediate state is only the vacuum one, i.e. $|0\rangle_{[\tilde{\Psi}^{\prime},\Omega^{\prime}]}{}_{[\tilde{\Psi}^{\prime},\Omega^{\prime}]}\langle 0|=|\text{sea}\rangle_{[\tilde{\Psi}^{\prime},\Omega^{\prime}]}{}_{[\tilde{\Psi}^{\prime},\Omega^{\prime}]}\langle \text{sea}|$. We may explicitly write the vacuum state (= Dirac sea) on a time-slice at the time $t$ as $|\text{sea}\rangle_{[\tilde{\Psi}^{\prime},\Omega^{\prime}](t)}$. Then the first line of (\ref{eq:2-identity}) reads 
\begin{align}
	{}_{[\tilde{\Psi}^{\prime},\Omega^{\prime}](+\infty)}
	\langle \text{sea}| 
	& Te^{-\int d^2\vec{x}\> H[\tilde{\Psi}^{\prime},g^{\prime}]}
	|\text{sea}\rangle_{[\tilde{\Psi}^{\prime},\Omega^{\prime}](-\infty)} 
	\nonumber \\
	& =\lim_{\genfrac{}{}{0pt}{}{t_F\to +\infty}{t_I\to -\infty}}
	{}_{[\tilde{\Psi}^{\prime},\Omega^{\prime}](t_F)}\langle \text{sea}|
	e^{-\int dx\> H[\tilde{\Psi}^{\prime},g^{\prime}]\Delta t}
	|\text{sea}
	\rangle_{[\tilde{\Psi}^{\prime},\Omega^{\prime}](t_F-\Delta t)} 
	\nonumber \\
	& \qquad \quad \> \> \times 
	{}_{[\tilde{\Psi}^{\prime},\Omega^{\prime}](t_F-\Delta t)}
	\langle \text{sea}|
	e^{-\int dx\> H[\tilde{\Psi}^{\prime},g^{\prime}]\Delta t}
	|\text{sea}
	\rangle_{[\tilde{\Psi}^{\prime},\Omega^{\prime}](t_F-2\Delta t)}
	\times \cdots \nonumber \\
	& \qquad \> \! \cdots 
	\times {}_{[\tilde{\Psi}^{\prime},\Omega^{\prime}](t_I+2\Delta t)}
	\langle \text{sea}|
	e^{-\int dx\> H[\tilde{\Psi}^{\prime},g^{\prime}]\Delta t}
	|\text{sea}
	\rangle_{[\tilde{\Psi}^{\prime},\Omega^{\prime}](t_I+\Delta t)} 
	\nonumber \\
	& \qquad \quad \> \> \times 
	{}_{[\tilde{\Psi}^{\prime},\Omega^{\prime}](t_I+\Delta t)}
	\langle \text{sea}|
	e^{-\int dx\> H[\tilde{\Psi}^{\prime},g^{\prime}]\Delta t}
	|\text{sea}\rangle_{[\tilde{\Psi}^{\prime},\Omega^{\prime}](t_I)} 
	\nonumber \\
	& \equiv \prod_{N=-\infty}^{+\infty}
	{}_{[\tilde{\Psi}^{\prime},\Omega^{\prime}](t_{N+1})}
	\langle \text{sea}|
	e^{-\int dx\> H[\tilde{\Psi}^{\prime},g^{\prime}]\Delta t}
	|\text{sea}\rangle_{[\tilde{\Psi}^{\prime},\Omega^{\prime}](t_N)}. 
	\tag{\ref{eq:2-identity}d}
	\label{eq:2-identity_d}
\end{align}
It should be noticed that the time axis $t$ is discretized such that 
\begin{align}
	t=t_N=N\Delta t,\quad (-\infty <N<+\infty ). 
	\label{eq:2-discretized_t}
\end{align}
Next the third line of (\ref{eq:2-identity}) becomes the amplitude on the metric 
\begin{align*}
	g_{\mu \nu}(\vec{x})=e^{2\Omega (\vec{x})}\eta_{\mu \nu}
\end{align*}
due to the Weyl transformation. Therefore we have to change the basis from $\left\{\frac{1}{\sqrt{2\pi}}\frac{e^{\frac{1}{2}(\Omega +\omega)}}{e^{n(t\pm ix)}}\right\}$ into $\left\{\frac{1}{\sqrt{2\pi}}\frac{e^{\frac{1}{2}\Omega}}{e^{n(t\pm ix)}}\right\}$. The reason of this change of the basis is because we would like to compare the creation and annihilation operators of the fields $\tilde{\Psi}^{\prime}$ and $\tilde{\Psi}$ which act to the same Hilbert space constructed from the operators expanded by the same basis $\left\{\frac{1}{\sqrt{2\pi}}\frac{e^{\frac{1}{2}\Omega}}{e^{n(t\pm ix)}}\right\}$. We then expand the field $\tilde{\Psi}^{\prime}$ in the two ways using two different above mentioned basis. 
\begin{align}
	\left\{ \begin{array}{l}
	\displaystyle \tilde{\psi}^{\prime}(\vec{x})
	=\frac{1}{\sqrt{2\pi}}\sum_{n=-\infty}^{+\infty}
	b_n^{[\tilde{\Psi}^{\prime},\Omega^{\prime}]}
	\frac{e^{\frac{1}{2}\Omega^{\prime}(\vec{x})}}{e^{n(t+ix)}}
	=\frac{1}{\sqrt{2\pi}}\sum_{n=-\infty}^{+\infty}
	b_n^{[\tilde{\Psi}^{\prime},\Omega]}
	\frac{e^{\frac{1}{2}\Omega (\vec{x})}}{e^{n(t+ix)}}, \\
	\displaystyle \tilde{\bar{\psi}}^{\prime}(\vec{x})
	=\frac{1}{\sqrt{2\pi}}\sum_{n=-\infty}^{+\infty}
	\bar{b}_n^{[\tilde{\Psi}^{\prime},\Omega^{\prime}]}
	\frac{e^{\frac{1}{2}\Omega^{\prime}(\vec{x})}}{e^{n(t-ix)}}
	=\frac{1}{\sqrt{2\pi}}\sum_{n=-\infty}^{+\infty}
	\bar{b}_n^{[\tilde{\Psi}^{\prime},\Omega]}
	\frac{e^{\frac{1}{2}\Omega (\vec{x})}}{e^{n(t-ix)}}. 
	\end{array} \right. 
	\label{eq:2-field.expansion_basis.trans}
\end{align}
Then subsequently we change the basis of the projection of the intermediate state $|\text{sea}\rangle_{[\tilde{\Psi}^{\prime},\Omega^{\prime}]}{}_{[\tilde{\Psi}^{\prime},\Omega^{\prime}]}\langle \text{sea}|$. As for Dirac sea we may write as 
\begin{align}
	|\text{sea}\rangle_{[\tilde{\Psi}^{\prime},\Omega^{\prime}]}
	& =\prod_{n\geq 0}b_n^{[\tilde{\Psi}^{\prime},\Omega^{\prime}]\dagger}
	|0\rangle \otimes 
	\prod_{m<0}\bar{b}_m^{[\tilde{\Psi}^{\prime},\Omega^{\prime}]\dagger}
	|\bar{0}\rangle \nonumber \\
	& 
	=\prod_{n\geq 0}\left(\int dx\> \tilde{\psi}^{\prime \dagger}(\vec{x})
	{}_{[\Omega^{\prime}]}\langle x|n\rangle 
	\right)|0\rangle \otimes 
	\prod_{m<0}\left(\int dx\> \tilde{\bar{\psi}}^{\prime \dagger}(\vec{x})
	{}_{[\Omega^{\prime}]}\langle x|\bar{m}\rangle \right)|\bar{0}\rangle 
	\nonumber \\
	& =\det_{n\geq 0}{}_{[\Omega^{\prime}]}\langle x|n\rangle \cdot 
	\det_{m<0}{}_{[\Omega^{\prime}]}\langle x|\bar{m}\rangle \times 
	\prod_x\tilde{\psi}^{\prime \dagger}(\vec{x})|0\rangle \otimes 
	\prod_x\tilde{\bar{\psi}}^{\prime \dagger}(\vec{x})|\bar{0}\rangle . 
	\tag{\ref{eq:2-identity}e}
	\label{eq:2-identity_e}
\end{align}
Now the projection operator can be written for $\vec{x}=(t,x)$ and $\vec{y}=(t,y)$ as 
\begin{align}
	|\text{sea}\rangle_{[\tilde{\Psi}^{\prime},\Omega^{\prime}]}
	{}_{[\tilde{\Psi}^{\prime},\Omega^{\prime}]}\langle \text{sea}|
	& =\det_{n\geq 0}{}_{[\Omega^{\prime}]}\langle x|n\rangle \cdot 
	\det_{m<0}{}_{[\Omega^{\prime}]}\langle x|\bar{m}\rangle \cdot 
	\det_{k\geq 0}\langle k|y\rangle_{[\Omega^{\prime}]} \cdot 
	\det_{l<0}\langle \bar{l}|y\rangle_{[\Omega^{\prime}]} \nonumber \\
	& \quad \times 
	\left\{\prod_x\tilde{\psi}^{\prime \dagger}(\vec{x})|0\rangle \otimes 
	\prod_x\tilde{\bar{\psi}}^{\prime \dagger}(\vec{x})|\bar{0}\rangle 
	\right\}
	\left\{\langle 0|\prod_y\tilde{\psi}^{\prime}(\vec{y})\otimes 
	\langle \bar{0}|\prod_y\tilde{\bar{\psi}}^{\prime}(\vec{y})
	\right\} \nonumber \\
	& 
	=\left\{\prod_x\tilde{\psi}^{\prime \dagger}(\vec{x})|0\rangle \otimes 
	\prod_x\tilde{\bar{\psi}}^{\prime \dagger}(\vec{x})|\bar{0}\rangle 
	\right\}
	\left\{\langle 0|\prod_y\tilde{\psi}^{\prime}(\vec{y})\otimes 
	\langle \bar{0}|\prod_y\tilde{\bar{\psi}}^{\prime}(\vec{y})
	\right\}. 
	\tag{\ref{eq:2-identity}f}
	\label{eq:2-identity_f}
\end{align}
Note that the expression (\ref{eq:2-identity_f}) is manifestly independent of the choice of the basis. Thus 
\begin{align}
	|\text{sea}\rangle_{[\tilde{\Psi}^{\prime},\Omega^{\prime}]}
	{}_{[\tilde{\Psi}^{\prime},\Omega^{\prime}]}\langle \text{sea}|
	& 
	=\left\{
	\prod_{n\geq 0}
	b_n^{[\tilde{\Psi}^{\prime},\Omega^{\prime}]\dagger}|0\rangle \otimes 
	\prod_{m<0}
	\bar{b}_m^{[\tilde{\Psi}^{\prime},\Omega^{\prime}]\dagger}
	|\bar{0}\rangle \right\}\left\{\langle 0|
	\prod_{n\geq 0}b_n^{[\tilde{\Psi}^{\prime},\Omega^{\prime}]}\otimes 
	\langle \bar{0}|
	\prod_{m<0}\bar{b}_m^{[\tilde{\Psi}^{\prime},\Omega^{\prime}]}
	\right\} \nonumber \\
	& 
	=\left\{\prod_{n\geq 0}
	b_n^{[\tilde{\Psi}^{\prime},\Omega]\dagger}|0\rangle \otimes 
	\prod_{m<0}
	\bar{b}_m^{[\tilde{\Psi}^{\prime},\Omega]\dagger}
	|\bar{0}\rangle \right\}\left\{\langle 0|\prod_{n\geq 0}
	b_n^{[\tilde{\Psi}^{\prime},\Omega]}\otimes 
	\langle \bar{0}|\prod_{m<0}
	\bar{b}_m^{[\tilde{\Psi}^{\prime},\Omega]}\right\} 
	\nonumber \\
	& \equiv |\text{sea}\rangle_{[\tilde{\Psi}^{\prime},\Omega]}
	{}_{[\tilde{\Psi}^{\prime},\Omega]}\langle \text{sea}|
	\tag{\ref{eq:2-identity}g}
	\label{eq:2-identity_g}
\end{align}
holds (Note the places of the prime). We may interpret $|\text{sea}\rangle_{[\tilde{\Psi}^{\prime},\Omega]}$ in the right hand side of (\ref{eq:2-identity_g}) as the Dirac sea on the curved space-time of the conformal factor $\Omega^{\prime}(\vec{x})$ in terms of the another space-time basis of the conformal factor $\Omega (\vec{x})$. In fact we may regard $|\text{sea}\rangle_{[\tilde{\Psi}^{\prime},\Omega]}$ as the pair production state in terms of the ``flat" (i.e. the metric $\Omega$) space-time.
\end{enumerate}

Here we summarize the notation of the Dirac seas in Table \ref{tab:notation}. We recall the Weyl transformation of the conformal component of the metric and the fermion field: 
\begin{align}
	& \Omega (\vec{x})\to 
	\Omega^{\prime}(\vec{x})=\Omega (\vec{x})+\omega (\vec{x}), 
	\nonumber \\
	& \tilde{\Psi}(\vec{x})\to 
	\tilde{\Psi}^{\prime}(\vec{x})=e^{\frac{1}{2}\omega (\vec{x})}
	\tilde{\Psi}(\vec{x}).
	\tag{\ref{eq:2-weyl.trans}}
\end{align}
The Dirac seas are defined by the various combinations of the two fermion fields and two metrics.
\begin{table}
\begin{center}
\extrarowheight=-4.5pt
\def\arraystretch{4}
\footnotesize
\begin{tabular}{|c|c|c|}\hline
	Dirac sea & operators and its expansions & view of field \\ \hline
	$\displaystyle |\text{sea}\rangle_{[\tilde{\Psi},\Omega]}
	=\prod_{n\geq 0}b_n^{[\tilde{\Psi},\Omega]\dagger}|0\rangle \otimes 
	\prod_{m<0}\bar{b}_m^{[\tilde{\Psi},\Omega]\dagger}|\bar{0}\rangle$ & 
	$\displaystyle b_n^{[\tilde{\Psi},\Omega]\dagger}
	=\frac{1}{\sqrt{2\pi}}\oint dx\> e^{-\Omega}\tilde{\psi}^{\dagger}
	\frac{e^{\frac{1}{2}\Omega}}{e^{n(t+ix)}}\>$ etc. & 
	$\displaystyle \tilde{\Psi}$ 
	on $g_{\mu \nu}=e^{2\Omega}\eta_{\mu \nu}$ \\ 
	\hline
	$\displaystyle 
	|\text{sea}\rangle_{[\tilde{\Psi}^{\prime},\Omega^{\prime}]}
	=\prod_{n\geq 0}b_n^{[\tilde{\Psi}^{\prime},\Omega^{\prime}]\dagger}
	|0\rangle \otimes 
	\prod_{m<0}\bar{b}_m^{[\tilde{\Psi}^{\prime},\Omega^{\prime}]\dagger}
	|\bar{0}\rangle$ & 
	$\displaystyle b_n^{[\tilde{\Psi}^{\prime},\Omega^{\prime}]\dagger}
	=\frac{1}{\sqrt{2\pi}}\oint dx\> 
	e^{-\Omega^{\prime}}\tilde{\psi}^{\prime \dagger}
	\frac{e^{\frac{1}{2}\Omega^{\prime}}}{e^{n(t+ix)}}\>$ etc. & 
	$\displaystyle \tilde{\Psi}^{\prime}$ on $g_{\mu \nu}^{\prime}
	=e^{2\Omega^{\prime}}\eta_{\mu \nu}$ \\ 
	\hline
	$\displaystyle |\text{sea}\rangle_{[\tilde{\Psi}^{\prime},\Omega]}
	=\prod_{n\geq 0}b_n^{[\tilde{\Psi}^{\prime},\Omega]\dagger}|0\rangle 
	\otimes 
	\prod_{m<0}\bar{b}_m^{[\tilde{\Psi}^{\prime},\Omega]\dagger}
	|\bar{0}\rangle$ & 
	$\displaystyle b_n^{[\tilde{\Psi}^{\prime},\Omega]\dagger}
	=\frac{1}{\sqrt{2\pi}}\oint dx\> 
	e^{-\Omega}\tilde{\psi}^{\prime \dagger}
	\frac{e^{\frac{1}{2}\Omega}}{e^{n(t+ix)}}\>$ etc. & 
	$\displaystyle \tilde{\Psi}^{\prime}$ on $g_{\mu \nu}
	=e^{2\Omega}\eta_{\mu \nu}$ \\ 
	\hline
\end{tabular}
	\caption{The notation of the various Dirac seas.}
	\label{tab:notation}
\end{center}
\end{table}
While we can define another Dirac sea $\displaystyle |\text{sea}\rangle_{[\tilde{\Psi},\Omega^{\prime}]}$, we don't need it in the current article.

\subsection{Pair production from Dirac sea corresponding to the Weyl transformation}\label{sec2.2}

In the present subsection, we calculate the transformation of the Dirac sea $|\text{sea}\rangle_{[\tilde{\Psi},\Omega](t)}$ on the time-slice $t$ under the Weyl transformation. Hereafter we assume that the parameter $\omega (\vec{x})$ in the Weyl transformation (\ref{eq:2-weyl.trans}) is infinitesimal. We now consider the Dirac sea $|\text{sea}\rangle_{[\tilde{\Psi}^{\prime},\Omega^{\prime}](t)}$ of the field $\tilde{\Psi}^{\prime}$ in the metric $g_{\mu \nu}^{\prime}=e^{2\Omega^{\prime}}\eta_{\mu \nu}=e^{2(\Omega +\omega )}\eta_{\mu \nu}$ and then we make a change of basis such that the state becomes $|\text{sea}\rangle_{[\tilde{\Psi}^{\prime},\Omega](t)}$ in the metric $g_{\mu \nu}=e^{2\Omega}\eta_{\mu \nu}$. In this case our concern is how the state $|\text{sea}\rangle_{[\tilde{\Psi}^{\prime},\Omega](t)}$ is expressed in terms of the creation and annihilation operators of the field $\tilde{\Psi}$ in the metric $g_{\mu \nu}=e^{2\Omega}\eta_{\mu \nu}$ (see eq. (\ref{eq:2-identity_g})). 

By making use of the infinitesimal Weyl transformation (\ref{eq:2-weyl.trans}), the amplitude in the third line of the identity (\ref{eq:2-identity}) may be obtained as follows. The fields $\tilde{\psi}^{\prime}$ and $\tilde{\bar{\psi}}^{\prime}$ should not be expanded by using the basis $\left\{\frac{1}{\sqrt{2\pi}}\frac{e^{\frac{1}{2}(\Omega +\omega )}}{e^{n(t\pm ix)}}\right\}$ but rather by using the basis $\left\{\frac{1}{\sqrt{2\pi}}\frac{e^{\frac{1}{2}\Omega}}{e^{n(t\pm ix)}}\right\}$, because in the third line the amplitude is the one in the metric $g_{\mu \nu}$ by the Weyl transformation. Thus 
\begin{align}
	\left\{ \begin{array}{l}
	\displaystyle \tilde{\psi}^{\prime}(\vec{x})
	=\frac{1}{\sqrt{2\pi}}\sum_{n=-\infty}^{+\infty}
	b_n^{[\tilde{\Psi}^{\prime},\Omega]}
	\frac{e^{\frac{1}{2}\Omega (\vec{x})}}{e^{n(t+ix)}}, \\
	\displaystyle \tilde{\bar{\psi}}^{\prime}(\vec{x})
	=\frac{1}{\sqrt{2\pi}}\sum_{n=-\infty}^{+\infty}
	\bar{b}_n^{[\tilde{\Psi}^{\prime},\Omega]}
	\frac{e^{\frac{1}{2}\Omega (\vec{x})}}{e^{n(t-ix)}}, 
	\end{array} \right. 
	\label{eq:2-weyled.field.expansion}
\end{align}
(see (\ref{eq:2-field.expansion_basis.trans})). We would like to express the expansion coefficients in (\ref{eq:2-weyled.field.expansion}), $b_n^{[\tilde{\Psi}^{\prime},\Omega]}$ and $\bar{b}_n^{[\tilde{\Psi}^{\prime},\Omega]}$ in terms of $b_n^{[\tilde{\Psi},\Omega]}$ and $\bar{b}_n^{[\tilde{\Psi},\Omega]}$ which become the physical creation and annihilation operators in the flat space-time as $t\to +\infty$. Here $b_n^{[\tilde{\Psi},\Omega]}$ and $\bar{b}_n^{[\tilde{\Psi},\Omega]}$ is obtained via 
\begin{align}
	\left\{ \begin{array}{l}
	\displaystyle \tilde{\psi}(\vec{x})
	=\frac{1}{\sqrt{2\pi}}\sum_{n=-\infty}^{+\infty}
	b_n^{[\tilde{\Psi},\Omega]}
	\frac{e^{\frac{1}{2}\Omega (\vec{x})}}{e^{n(t+ix)}}, \\
	\displaystyle \tilde{\bar{\psi}}(\vec{x})
	=\frac{1}{\sqrt{2\pi}}\sum_{n=-\infty}^{+\infty}
	\bar{b}_n^{[\tilde{\Psi},\Omega]}
	\frac{e^{\frac{1}{2}\Omega (\vec{x})}}{e^{n(t-ix)}}. 
	\end{array} \right. 
	\label{eq:2-field.expansion}
\end{align}
The change of the basis and thus the change of the expressions of the creation and annihilation operators remind us the Bogolyubov transformation. Then we may expect the appearance of $J_{\omega}$ in the third line of (\ref{eq:2-identity}). We will show below that there appears pair production states from the Dirac sea. 

\begin{figure}[htb]
\begin{tabular}{cc}
\begin{minipage}{0.48\textwidth}
\begin{center}
\includegraphics[width=55mm,angle=270]{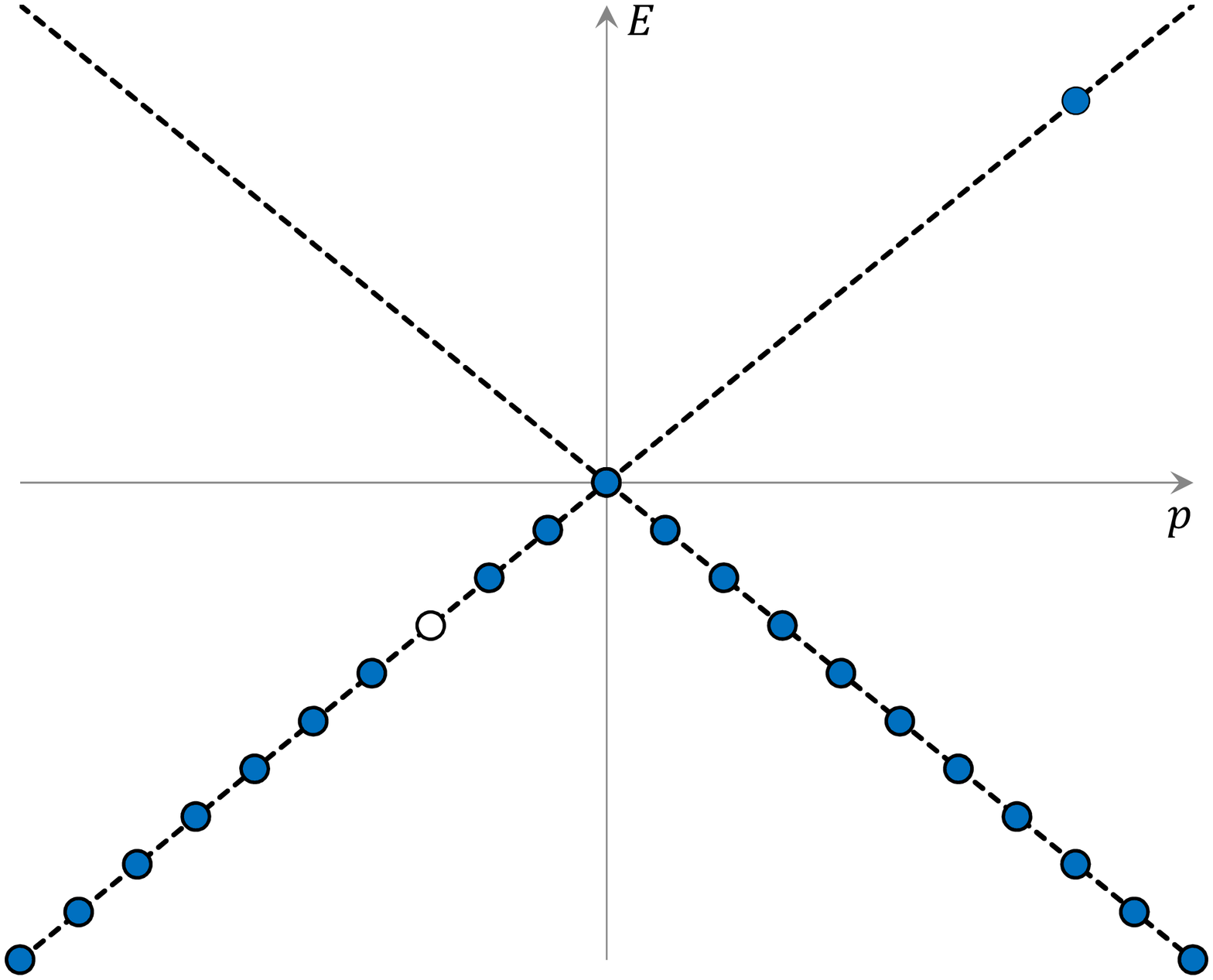}
\par{$\prod_{n\geq 0}b_n^{[\tilde{\Psi},\Omega]\dagger}|0\rangle \otimes |\bar{8},\bar{3}\rangle_{[\tilde{\Psi},\Omega]}$}
\end{center}
\end{minipage} & 
\begin{minipage}{0.48\textwidth}
\begin{center}
\includegraphics[width=55mm,angle=270]{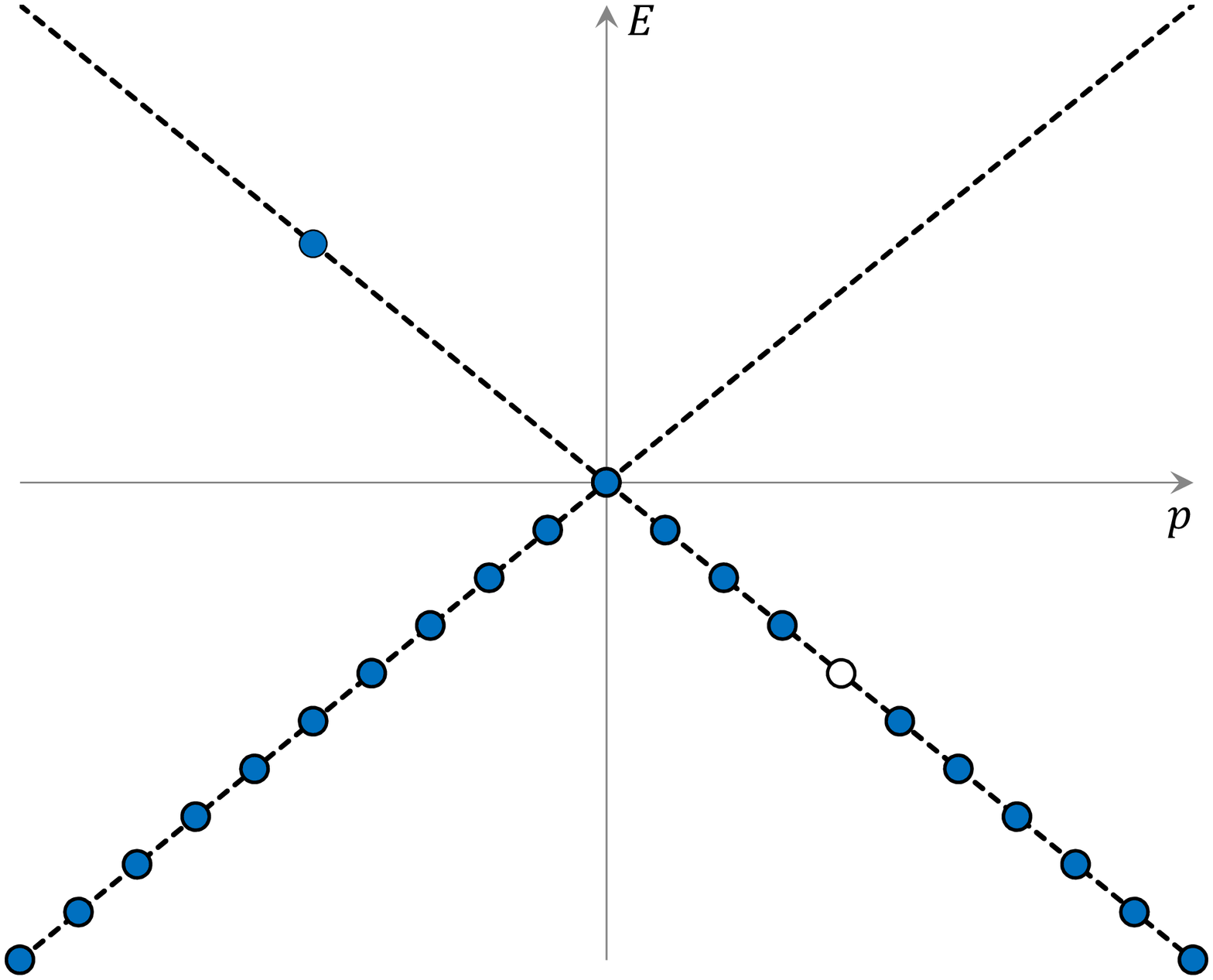}
\par{$|5,4\rangle_{[\tilde{\Psi},\Omega]}\otimes \prod_{\bar{n}<0}\bar{b}_{\bar{n}}^{[\tilde{\Psi},\Omega]\dagger}|\bar{0}\rangle$}
\end{center}
\end{minipage}
\end{tabular}
\caption[Pair production states]{The examples of the pair production states. The white circles in the lower half planes represent the holes.}
\label{pic:pair.prod}
\end{figure}

In the first step we derive the relations of the two expansion coefficients $b_n^{[\tilde{\Psi}^{\prime},\Omega]}$ and $b_n^{[\tilde{\Psi},\Omega]}$. We shoud notice that the coefficients $b_n^{[\tilde{\Psi}^{\prime},\Omega]}$ are the components of the right hand side of (\ref{eq:2-identity_g}). From (\ref{eq:2-weyl.trans}), (\ref{eq:2-weyled.field.expansion}) and (\ref{eq:2-field.expansion}), we find 
\begin{align}
	b_n^{[\tilde{\Psi}^{\prime},\Omega]}
	& =\oint \frac{dx}{2\pi}\> e^{-\Omega (\vec{x})}\cdot 
	e^{\frac{1}{2}\Omega (\vec{x})}e^{n(t+ix)}
	\left\{1+\frac{1}{2}\omega (\vec{x})\right\}\sum_{m=-\infty}^{+\infty}
	b_m^{[\tilde{\Psi},\Omega]}
	\frac{e^{\frac{1}{2}\Omega (\vec{x})}}{e^{m(t+ix)}} \nonumber \\
	& \equiv \sum_{m=-\infty}^{+\infty}W_{n,m}b_m^{[\tilde{\Psi},\Omega]}, 
	\nonumber \\
	\bar{b}_n^{[\tilde{\Psi}^{\prime},\Omega]}
	& =\oint \frac{dx}{2\pi}\> e^{-\Omega (\vec{x})}\cdot 
	e^{\frac{1}{2}\Omega (\vec{x})}e^{n(t-ix)}
	\left\{1+\frac{1}{2}\omega (\vec{x})\right\}\sum_{m=-\infty}^{+\infty}
	\bar{b}_m^{[\tilde{\Psi},\Omega]}
	\frac{e^{\frac{1}{2}\Omega (\vec{x})}}{e^{m(t-ix)}} \nonumber \\
	& \equiv \sum_{m=-\infty}^{+\infty}\bar{W}_{n,m}
	\bar{b}_m^{[\tilde{\Psi},\Omega]}. 
	\label{eq:2-weyl.trans-relation}
\end{align}
Here we defined a matrix representation of the infinitesimal Weyl transformation which is time dependent as 
\begin{align}
	W_{n,m}(t)
	& \equiv \delta_{n,m}+\int dx\> 
	\left\{\frac{1}{2}\omega (\vec{x})\right\}
	\langle n|x\rangle_{[\Omega]}{}_{[\Omega]}\langle x|m\rangle 
	=\delta_{n,m}+\oint \frac{dx}{2\pi}\> 
	\left\{\frac{1}{2}\omega (\vec{x})\right\}e^{(n-m)(t+ix)}, \nonumber \\
	\bar{W}_{n,m}(t)
	& \equiv \delta_{n,m}+\int dx\> 
	\left\{\frac{1}{2}\omega (\vec{x})\right\}
	\langle \bar{n}|x\rangle_{[\Omega]}
	{}_{[\Omega]}\langle x|\bar{m}\rangle 
	=\delta_{n,m}+\oint \frac{dx}{2\pi}\> 
	\left\{\frac{1}{2}\omega (\vec{x})\right\}e^{(n-m)(t-ix)}.
	\label{eq:2-weyl.trans-matrix}
\end{align}
We find that the commutation relations for $b_n^{[\tilde{\Psi}^{\prime},\Omega]}$ and $\bar{b}_n^{[\tilde{\Psi}^{\prime},\Omega]}$ are not the ordinary ones: 
\begin{align*}
	\Big\{b_n^{[\tilde{\Psi}^{\prime},\Omega]}, & 
	b_m^{[\tilde{\Psi}^{\prime},\Omega]\dagger}\Big\} \\
	& =\left\{\sum_k W_{n,k}b_k^{[\tilde{\Psi},\Omega]}, 
	\sum_l W_{m,l}^{\dagger}b_l^{[\tilde{\Psi},\Omega]\dagger}\right\}
	=\sum_{k=-\infty}^{+\infty}W_{n,k}W_{m,k}^{\dagger} \\
	& =\sum_{k=-\infty}^{+\infty}
	\left[\delta_{n,k}+\oint \frac{dx}{2\pi}\> 
	\left\{\frac{1}{2}\omega (\vec{x})\right\}e^{(n-k)(t+ix)}\right]
	\left[\delta_{m,k}+\oint \frac{dx}{2\pi}\> 
	\left\{\frac{1}{2}\omega (\vec{x})\right\}e^{-(m-k)(t+ix)}\right] \\
	& =\delta_{n,m}+\oint \frac{dx}{2\pi}\> \omega (\vec{x})e^{(n-m)(t+ix)}
\end{align*}
The similar relation also holds for $\bar{b}_n^{[\tilde{\Psi}^{\prime},\Omega]}$. This $\omega$ correction to the algebra of the harmonic oscillator may cause the Weyl anomaly. In terms of these representation of the Weyl transformation the Dirac sea of the right hand side of (\ref{eq:2-identity_g}) reads up to the first order of $\omega$ 
\begin{align}
	|\text{sea}\rangle_{[\tilde{\Psi}^{\prime},\Omega](t)}
	& =\prod_{n\geq 0}b_n^{[\tilde{\Psi}^{\prime},\Omega]\dagger}
	|0\rangle \otimes 
	\prod_{\bar{n}<0}
	\bar{b}_{\bar{n}}^{[\tilde{\Psi}^{\prime},\Omega]\dagger}
	|\bar{0}\rangle \nonumber \\
	& =\prod_{n\geq 0}\left(\sum_{m=-\infty}^{+\infty}W_{n,m}^{\dagger}
	b_m^{[\tilde{\Psi},\Omega]\dagger}\right)|0\rangle \otimes 
	\prod_{\bar{n}<0}\left(\sum_{\bar{m}=-\infty}^{+\infty}
	\bar{W}_{\bar{n},\bar{m}}^{\dagger}
	\bar{b}_{\bar{m}}^{[\tilde{\Psi},\Omega]\dagger}\right)
	|\bar{0}\rangle \nonumber \\
	& =\det \left[\delta_{n,m}+\oint \frac{dx}{2\pi}\> 
	\left\{\frac{1}{2}\omega (\vec{x})\right\}
	e^{-(n-m)(t+ix)}\right]_{n,m\geq 0} \nonumber \\
	& \qquad \times \det \left[\delta_{\bar{n},\bar{m}}
	+\oint \frac{dx}{2\pi}\> 
	\left\{\frac{1}{2}\omega (\vec{x})\right\}
	e^{(\bar{n}-\bar{m})(t-ix)}\right]_{\bar{n},\bar{m}>0}|
	\text{sea}\rangle_{[\tilde{\Psi},\Omega](t)} \nonumber \\
	& \quad +\sum_{n\geq 0,m>0}\oint \frac{dx}{2\pi}\> 
	\left\{\frac{1}{2}\omega (\vec{x})\right\}e^{-(n+m)(t+ix)}
	|m,n\rangle_{[\tilde{\Psi},\Omega](t)}
	\otimes \prod_{\bar{n}<0}
	\bar{b}_{\bar{n}}^{[\tilde{\Psi},\Omega]\dagger}
	|\bar{0}\rangle \nonumber \\
	& \quad +\sum_{\bar{n}>0,\bar{m}\geq 0}\oint \frac{dx}{2\pi}\> 
	\left\{\frac{1}{2}\omega (\vec{x})\right\}e^{(\bar{n}+\bar{m})(t-ix)}
	\prod_{n\geq 0}b_n^{[\tilde{\Psi},\Omega]\dagger}|0\rangle 
	\otimes |\bar{m},\bar{n}\rangle_{[\tilde{\Psi},\Omega](t)}.
	\label{eq:2-weyl.dirac.a}
\end{align}
The reason why the factor in front of $|\text{sea}\rangle_{[\tilde{\Psi},\Omega](t)}$ in the third and fourth lines of (\ref{eq:2-weyl.dirac.a}) becomes $\det W_{n,m}^{\dagger}\times \det \bar{W}_{\bar{n},\bar{m}}^{\dagger}$ is presented in Appendix C. In this expression we notice that $|\text{sea}\rangle_{[\tilde{\Psi}^{\prime},\Omega](t)}$ is a superposition as a sum of the three types of states: 
\begin{align*}
	& |\text{sea}\rangle_{[\tilde{\Psi},\Omega](t)}, \\
	& |m,n\rangle_{[\tilde{\Psi},\Omega](t)}\otimes \prod_{\bar{n}<0}
	\bar{b}_{\bar{n}}^{[\tilde{\Psi},\Omega]\dagger}|\bar{0}\rangle 
\end{align*}
and 
\begin{align*}
	\prod_{n\geq 0}b_n^{[\tilde{\Psi},\Omega]\dagger}|0\rangle \otimes 
	|\bar{m},\bar{n}\rangle_{[\tilde{\Psi},\Omega](t)}. 
\end{align*}
Here $|m,n\rangle_{[\tilde{\Psi},\Omega](t)}$ and $|\bar{m},\bar{n}\rangle_{[\tilde{\Psi},\Omega](t)}$ are defined by 
\begin{align}
	|m,n\rangle_{[\tilde{\Psi},\Omega](t)} 
	& \equiv b_0^{[\tilde{\Psi},\Omega]\dagger}
	b_1^{[\tilde{\Psi},\Omega]\dagger}\cdots 
	b_{n-1}^{[\tilde{\Psi},\Omega]\dagger}
	b_{-m}^{[\tilde{\Psi},\Omega]\dagger}
	b_{n+1}^{[\tilde{\Psi},\Omega]\dagger}\cdots |0\rangle , 
	\quad \text{ for }n\geq 0,m>0, \nonumber \\
	|\bar{m},\bar{n}\rangle_{[\tilde{\Psi},\Omega](t)} 
	& \equiv 
	\bar{b}_{-\bar{1}}^{[\tilde{\Psi},\Omega]\dagger}
	\bar{b}_{-\bar{2}}^{[\tilde{\Psi},\Omega]\dagger}\cdots 
	\bar{b}_{-\bar{n}+1}^{[\tilde{\Psi},\Omega]\dagger}
	\bar{b}_{\bar{m}}^{[\tilde{\Psi},\Omega]\dagger}
	\bar{b}_{-\bar{n}-1}^{[\tilde{\Psi},\Omega]\dagger}\cdots 
	|\bar{0}\rangle , \quad \text{ for }\bar{n}>0,\bar{m}\geq 0. 
	\label{eq:2-pair.prod}
\end{align}
Let us investigate in detail of the equation (\ref{eq:2-weyl.dirac.a}). In the terms involving $b_{m\geq 0}^{[\tilde{\Psi},\Omega]\dagger}$ and $\bar{b}_{m<0}^{[\tilde{\Psi},\Omega]\dagger}$ and summing up over $m$ and $\bar{m}$ in the second line of (\ref{eq:2-weyl.dirac.a}), the contribution coming from the representation matrix (\ref{eq:2-weyl.trans-matrix}) of the Weyl transformation turns out to become the determinant and as the state it is united to the Dirac sea $|\text{sea}\rangle_{[\tilde{\Psi},\Omega](t)}$ as the third and fourth lines of (\ref{eq:2-weyl.dirac.a}). As for the terms containing $b_{m<0}^{[\tilde{\Psi},\Omega]\dagger}$ in the second line of (\ref{eq:2-weyl.dirac.a}), they create the pair-produced states, the left mover particles with energy $m$ and left mover holes (= anti-particles) with energy $n$ that can be written as 
\begin{align*}
	|m,n\rangle_{[\tilde{\Psi},\Omega](t)}\otimes \prod_{\bar{n}<0}
	\bar{b}_{\bar{n}}^{[\tilde{\Psi},\Omega]\dagger}|\bar{0}\rangle . 
\end{align*}
At the same time the terms containing $\bar{b}_{m\geq 0}^{[\tilde{\Psi},\Omega]\dagger}$ creates the pair-produced states, the right mover particles with energy $\bar{m}$ and the right mover holes (= anti-particles) with energy $\bar{n}$, which may be written as 
\begin{align*}
	\prod_{n\geq 0}b_n^{[\tilde{\Psi},\Omega]\dagger}|0\rangle \otimes 
	|\bar{m},\bar{n}\rangle_{[\tilde{\Psi},\Omega](t)}. 
\end{align*}
In summary, the second line of (\ref{eq:2-weyl.dirac.a}) is reduced to be a linear combination of the Dirac sea and the pair production states expressed as the last four lines of (\ref{eq:2-weyl.dirac.a}). Some simple examples of the dispersion relations of such pair production states are depicted in Figure~\ref{pic:pair.prod}. 
From now on, the pair production states are represented as 
\begin{align*}
	|\text{pair}\rangle_{[\tilde{\Psi},\Omega](t)}
\end{align*}
in short. Strictly speaking, we can use the word ``pair production" only when $\Omega =0$, i.e. in the flat space-time. Here, we use it symbolically but it expresses the point. By applying the formula $\det e^U=e^{\tr U}$ of a matrix $U$ to (\ref{eq:2-weyl.dirac.a}), we obtain 
\begin{align}
	|\text{sea} & \rangle_{[\tilde{\Psi}^{\prime},\Omega](t)} \nonumber \\
	& =\exp \left\{\sum_{n\! \text{ or }\! \bar{n}\geq 0}\int dx\> 
	\frac{1}{2}\omega (\vec{x})
	\Big(\langle n|x\rangle_{[\Omega](t)}
	{}_{[\Omega](t)}\langle x|n\rangle 
	+\langle \bar{n}|x\rangle_{[\Omega](t)}
	{}_{[\Omega](t)}\langle x|\bar{n}\rangle 
	\Big)\right\}|\text{sea}\rangle_{[\tilde{\Psi},\Omega](t)} \nonumber \\
	& \quad +|\text{pair}\rangle_{[\tilde{\Psi},\Omega](t)} 
	\label{eq:2-weyl.dirac.c}
\end{align}
Obviously the ket vector (\ref{eq:2-weyl.dirac.c}) belongs to the Fock space $\mathcal{H}_{\Omega}$ and its hermitian conjugation should be taken in terms of the inner product of $\mathcal{H}_{\Omega}$. The expression (\ref{eq:2-weyl.dirac.c}) is the Dirac sea $|\text{sea}\rangle_{[\tilde{\Psi}^{\prime},\Omega](t)}$ at the time-slice $t$ arising from the Weyl transformation and the factor in front of $|\text{sea}\rangle_{[\tilde{\Psi},\Omega](t)}$ will turn out to be a Weyl anomaly. Such a Dirac sea stems from pair production of the particles. Strictly speaking the pair production state $|\text{pair}\rangle_{[\tilde{\Psi},\Omega](t)}$ does not contribute when taking the expectation value at each time interval of the amplitude (\ref{eq:2-identity_g}) in the first order of $\omega$ and thus the pair production state does not directly contribute to the Weyl anomaly. However by considering (\ref{eq:2-weyl.dirac.c}), the pair production state $|\text{pair}\rangle_{[\tilde{\Psi},\Omega](t)}$ takes some norm from $|\text{sea}\rangle_{[\tilde{\Psi}^{\prime},\Omega](t)}$ and create the state $|\text{sea}\rangle_{[\tilde{\Psi},\Omega](t)}$ which contributes directly to the Weyl anomaly. In this sense we may say that the pair production state is the origin of the Weyl anomaly. Furthermore if we take into account of the higher order terms of $\omega$, the state including states more than two pairs production may directly contribute to the Weyl anomaly. 

The Weyl anomaly is produced by summing up through all history from the infinite past to infinite future as (\ref{eq:2-identity_d}) the exponential factor appearing in front of the Dirac sea $|\text{sea}\rangle_{[\tilde{\Psi},\Omega](t)}$ at each time-slice $t=t_N$ of (\ref{eq:2-weyl.dirac.c}). In (\ref{eq:2-weyl.dirac.c}) the summation $\sum_{n\! \text{ or }\! \bar{n}\geq 0}$ may diverge and its divergence corresponds to the ultra-violet one of the space coordinate $x$. On the other hand the product $\prod_N$ in (\ref{eq:2-identity_d}) may also diverge. As we will see in (\ref{eq:2-weyl.dirac.c''_temp}), this divergence turns out to be that of the integration over the ``energy" $k^0$, i.e. the ultra-violet divergence of the time coordinate $t$. We should notice that this divergence is natural and plausible because we adopt the general covariant theory. In the next subsection we shall indeed perform the derivation of the Weyl anomaly in the above method.

\subsection{Evaluation of $J_{\omega}$ and the Weyl anomaly}\label{sec2.3}

In this subsection we compute $J_{\omega}$ in (\ref{eq:2-identity}) by the method described in the last part of the previous subsection B. Firstly we divide the time direction into an infinitesimal time interval $t_N$ and at each time in the amplitude (\ref{eq:2-identity_d}) we apply the changing formula of the basis (\ref{eq:2-identity_g}) of the projections. As is described there, $J_{\omega}$ is obtained by summing up over all time the contribution (\ref{eq:2-weyl.dirac.c}) and its hermitian conjugation coming from the Weyl transformation of the intermediate Dirac sea at each time-slice. Then $J_{\omega}$ reads 
\begin{align}
	J_{\omega}
	& =\prod_{N=-\infty}^{+\infty}
	\exp \left\{2\times \sum_{n\! \text{ or }\! \bar{n}\geq 0}\int dx\> 
	\frac{1}{2}\omega (t_N,x)\Big(
	\langle n|x\rangle_{[\Omega](t_N)}{}_{[\Omega](t_N)}\langle x|n\rangle 
	+\langle \bar{n}|x\rangle_{[\Omega](t_N)}{}_{[\Omega](t_N)}\langle x|
	\bar{n}\rangle \Big)\right\}. 
	\label{eq:2-weyl.dirac.c'}
\end{align}
Here $t_N$ is defined in (\ref{eq:2-discretized_t}). The reader should remember that the product $\prod_N$ is taken by the step $\Delta t=t_N-t_{N-1}$. As we show explicitly in eq. (\ref{eq:2-weyl.dirac.c''_temp}), this product with respect to $N$ is exponentiated and becomes a summation $\sum_N$. We furthermore take the continuum limit, $\Delta t\to 0$, and it becomes the integration over the time $t$, $\int dt$. It should be also noted that the overall factor $2$ in the exponent comes from the fact that the bra and ket vectors both contribute. We should be aware the fact that (\ref{eq:2-weyl.dirac.c'}) is non-unitary transformation because of the reality of the exponent which is the consequence of the reality of the matrix $W_{n,m}$ in (\ref{eq:2-weyl.trans-matrix}). This is rather naturally interpreted similar to the case of Schwinger mechanism \cite{rf:schwinger} and Hawking radiation, because the non-unitarity generates the pair production. On the other hand, as will be shown in Appendix B, the conformal transformation does not lead to pair production since the contribution by the transformation is unitary (the exponent is pure imaginary).

In the course of the calculation of $J_{\omega}$, we face the divergence problem in the sense of $\sum_{n=0}^{\infty}1$. We would like to use the heat kernel regularization to tame this divergence.

We now consider the differential operator in the action (\ref{eq:2-rescaled.action}) 
\begin{align}
	& \Slash{D}\equiv e^{-\frac{1}{2}\Omega (\vec{x})}\gamma^0\gamma^i
	\partial_ie^{-\frac{1}{2}\Omega (\vec{x})}
	\label{eq:2-hermitian.operator}
\end{align}
and solve the eigenvalue problem 
\begin{align}
	& \Slash{D}^{\dagger}\Slash{D}\chi_r(\vec{x})
	=\lambda_r^2\chi_r(\vec{x}) 
	\label{eq:2-heat.kernel}
\end{align}
with the eigenvalues $\lambda_r^2$ and a set of complete orthonormal eigenfunctions $\left\{\chi_r(\vec{x})\right\}$. Then the product $\prod_N$ appearing in (\ref{eq:2-weyl.dirac.c'}), we exponentiate to make it the summation over $N$. Thus after all it becomes the trace of each time $t$, 
\begin{align}
	J_{\omega}
	& =\exp \left\{\sum_{n\geq 0}\sum_{N=-\infty}^{+\infty}\int dx\> 
	\omega (t_N,x)\Big(
	\langle n|x\rangle_{[\Omega](t_N)}{}_{[\Omega](t_N)}\langle x|n\rangle 
	+\langle \bar{n}|x\rangle_{[\Omega](t_N)}{}_{[\Omega](t_N)}\langle x|
	\bar{n}\rangle \Big)\right\} \nonumber \\
	& =\exp \Bigg\{\sum_{n\geq 0}\sum_{N=-\infty}^{+\infty}\int dx\> 
	\int_{-\frac{\pi}{\Delta t}}^{+\frac{\pi}{\Delta t}}dk^0
	\sqrt{\frac{\Delta t}{2\pi}}e^{ik^0t_N}\omega (t_N,x) \nonumber \\
	& \qquad \qquad \qquad \qquad \qquad \times \Big(
	{}_{[\Omega](t_N)}\langle x|n\rangle \langle n|x\rangle_{[\Omega](t_N)}
	+{}_{[\Omega](t_N)}\langle x|\bar{n}\rangle \langle \bar{n}|
	x\rangle_{[\Omega](t_N)}\Big)
	\sqrt{\frac{\Delta t}{2\pi}}e^{-ik^0t_N}\Bigg\}
	\label{eq:2-weyl.dirac.c''_temp}
\end{align}
Here $\left\{\sqrt{\frac{\Delta t}{2\pi}}e^{-ik^0t_N}\right\}$ denotes the basis of the plane wave with respect to the discretized time $t_N=N\Delta t$ and $\left\{\frac{1}{\sqrt{2\pi}}e^{-ik^0t}\right\}$ is the continuum limit $\Delta t\to 0$ of it\footnote{Formally the summation $\sum_N\cdots$ means the trace $\sum_N \langle t_N|\cdots |t_N\rangle$. However according to the Pauli's argument the eigenstate of the time $|t\rangle$ cannot exist in our case of the present article. Therefore we use the plane wave $e^{-ik_0t_N}$ instead of $|t_N\rangle$. If we treat the Dirac sea not in $2$-dimensional but in $4$-dimensional space-time, $|t\rangle$ can exist since the energy spectrum is continuous and unbounded. The ``normalization" of the basis $\left\{\sqrt{\Delta t}e^{-ik^0t_N}\right\}$ is written as $\langle k_0|k_0\rangle =\sum_N \sqrt{\Delta t}e^{ik_0t_N}\cdot \sqrt{\Delta t}e^{-ik_0t_N}=\sum_N \Delta t$. When we take the limit $\Delta t\to 0$, it becomes $\langle k_0|k_0\rangle =\int dt\> e^{ik_0t}\cdot e^{-ik_0t}=\int dt$, i.e. the volume of the time.}. We take the continuum limit $\Delta t\to 0$, and the summation $\sum_N $ and the time step $\Delta t$ join to the integration over $t$, $\int dt$. Then the exponent of (\ref{eq:2-weyl.dirac.c''_temp}) becomes the integration over the whole space-time, 
\begin{align}
	J_{\omega}
	& =\exp \left\{\sum_{n\geq 0}\int \frac{dk^0}{2\pi}\int d^2\vec{x}\> 
	\omega (\vec{x})e^{ik^0t}\Big({}_{[\Omega](t)}\langle x|n\rangle 
	\langle n|x\rangle_{[\Omega](t)}+
	{}_{[\Omega](t)}\langle x|\bar{n}\rangle 
	\langle \bar{n}|x\rangle_{[\Omega](t)}\Big)e^{-ik^0t}\right\} 
	\nonumber \\
	& =\exp \left\{\int \frac{dk^0}{2\pi}\int d^2\vec{x}\> \omega (\vec{x})
	e^{ik^0t}\Big({}_{[\Omega](t)}\langle x|x\rangle_{[\Omega](t)}\Big)
	e^{-ik^0t}\right\} \nonumber \\
	& =\exp \left\{\int \frac{d^2\vec{k}}{(2\pi )^2}\int d^2\vec{x}\> 
	\omega (\vec{x})e^{i\vec{k}\cdot \vec{x}}e^{-i\vec{k}\cdot \vec{x}}
	\right\}. 
	\label{eq:2-weyl.dirac.c''}
\end{align}
Note that it is here understood that the $dk^0$-integration is cut off at $\pm \frac{\pi}{\Delta t}$ as can be seen from (\ref{eq:2-weyl.dirac.c''_temp}) explicitly. This means that $\Delta t$ is really representing an ultraviolet cut-off. This cut-off interpretation of $\Delta t$ also explains the mysterious appearance of (\ref{eq:2-weyl.dirac.c'}) in which the various factors --the exponentials-- contain no compensating dependence on the stepping distance in time $\Delta t=t_N-t_{N-1}$, so that $J_{\omega}$ gets strongly depending on this stepping distance $\Delta t$ (Interpreted as cut-off this strange dependence may be O.K.). The kets $|n\rangle$ and $|\bar{n}\rangle$ are the complete set in $-\infty <n<+\infty$. In fact the quantities of $(\cdots )$ in the exponent of the first line of (\ref{eq:2-weyl.dirac.c''}) is independent of $n$, because ${}_{[\Omega]}\langle x|n\rangle$ is just a plane wave essentially. Thus we may rewrite $\sum_{n\geq 0}=\frac{1}{2}\sum_{n=-\infty}^{+\infty}$ and use there the completeness relation $\sum_n|n\rangle \langle n|=\sum_n|\bar{n}\rangle \langle \bar{n}|=1$. By changing the basis from $\left\{e^{-i\vec{k}\cdot \vec{x}}\right\}$ to $\left\{\chi_r(\vec{x})\right\}$, $J_{\omega}$ in (\ref{eq:2-weyl.dirac.c''}) reads 
\begin{align}
	J_{\omega}
	& =\exp \left[\int d^2\vec{x} \> \omega (\vec{x})
	\sum_{r=0}^{+\infty}\chi_r^{\dagger}(\vec{x})\chi_r(\vec{x})\right]. 
	\label{eq:2-weyl.dirac.d}
\end{align}
We then next introduce the exponential regulator of $\lambda_r$ that involves the regularization cut-off $M$. The divergent quantity $J_{\omega}$ in (\ref{eq:2-weyl.dirac.d}) is now redefined as 
\begin{align}
	J_{\omega}
	& \equiv \exp \left[\lim_{M\to \infty}\int d^2\vec{x} \> 
	\omega (\vec{x})\sum_{r=0}^{+\infty}\chi_r^{\dagger}(\vec{x})
	e^{-\frac{\lambda_r^2}{M^2}}\chi_r(\vec{x})\right]. 
	\label{eq:2-weyl.dirac.e}
\end{align}
Finally by expanding in powers of the cut-off $M$\footnote{As for the calculation see e.g. Ref.~\cite{rf:fujikawa}.} and inserting $\sqrt{g}=e^{2\Omega (\vec{x})}$ and $\sqrt{g}R=-2\partial_{\mu}\partial^{\mu}\Omega (\vec{x})$ we obtain 
\begin{align}
	J_{\omega}
	& =\exp \left[\frac{1}{2\pi}\lim_{M\to \infty}\int d^2\vec{x} \> 
	\omega (\vec{x})\left\{M^2e^{2\Omega (\vec{x})}
	-\frac{1}{12}\cdot 2\partial_{\mu}\partial^{\mu}\Omega (\vec{x})
	\right\}\right] \nonumber \\
	& =\exp \left[\frac{1}{2\pi}\lim_{M\to \infty}\int d^2\vec{x} \> 
	\omega (\vec{x})
	\left\{M^2\sqrt{g}+\frac{1}{12}\sqrt{g}R\right\}\right]. 
	\label{eq:2-weyl.dirac}
\end{align}
The finite part of (\ref{eq:2-weyl.dirac}) is proportional to the exponentiated Ricci scalar and we may expect that it is the Weyl anomaly. However, we need the relation with the trace $T^{\mu}_{\> \mu}$ of the energy-momentum tensor. Indeed we relate the Ricci scalar in the exponent of (\ref{eq:2-weyl.dirac}) and $T^{\mu}_{\> \mu}$ by using the second and third lines in the identity (\ref{eq:2-identity}). 

We in fact derive the Weyl anomaly by using the second and third lines of the identity (\ref{eq:2-identity}), 
\begin{align}
	\langle \text{sea}|T\exp \left(-\int d^2\vec{x}\> 
	H[\tilde{\Psi},g^{\prime}]\right)|\text{sea}\rangle 
	=\langle \text{sea}|J_{\omega}\cdot 
	T\exp \left(-\int d^2\vec{x}\> H[\tilde{\Psi},g]\right)
	|\text{sea}\rangle . 
	\label{eq:2-anomaly.identity}
\end{align}
Here the limit 
\begin{align*}
	\lim_{t\to \pm \infty}\Omega (\vec{x})
	=\lim_{t\to \pm \infty}\omega (\vec{x})=0
\end{align*}
is taken. Thus the Dirac sea of the initial and final states in (\ref{eq:2-anomaly.identity}) are those on the flat space-time 
\begin{align*}
	|\text{sea}\rangle 
	\equiv |\text{sea}\rangle _{[\tilde{\Psi},\Omega =0](t=\pm \infty)}. 
\end{align*}
The $J_{\omega}$ is obtained by renormalizing the divergent term proportional to $M^2$ in (\ref{eq:2-weyl.dirac}) into the cosmological constant term, 
\begin{align}
	J_{\omega}=\exp \left[\int d^2\vec{x}\> 
	\omega (\vec{x})\left\{\frac{1}{24\pi}\sqrt{g}R\right\}\right]. 
	\label{eq:2-jacobian}
\end{align}
Remembering the argument in the subsection \ref{sec2.1} the terms in the left and right hands sides are both obtained by integrating the same field $\tilde{\Psi}$. However we rewrite them by using different metrics to produce the different effective actions $S_{\text{eff}}[g=e^{2\Omega}]=\int dt\> E_0^{[\tilde{\Psi},\Omega]}$, 
\begin{align}
	e^{-S_{\text{eff}}[g^{\prime}=e^{2(\Omega +\omega )}]}
	=J_{\omega}\cdot e^{-S_{\text{eff}}[g=e^{2\Omega}]}. 
\end{align}
By taking logarithm on both sides, the difference of the effective action $S_{\text{eff}}$ becomes to the differentiation with respect to the infinitesimal $\omega$, 
\begin{align}
	\int d^2\vec{x}\> \langle -2\omega T^{\mu}_{\> \mu}\rangle 
	=\int d^2\vec{x}\> \omega \left\{\frac{1}{24\pi}\sqrt{g}R\right\}. 
\end{align}
This is the equality what we are looking for, i.e. the relation of the Weyl anomaly: 
\begin{align}
	\langle T^{\mu}_{\> \mu}\rangle =-\frac{1}{48\pi}\sqrt{g}R. 
	\label{eq:2-final.result}
\end{align}
%

\section{Conclusion and outlooks}\label{sec3}

\subsection{Conclusion}\label{sec3.1}

In the present article we presented that when the Dirac sea in the background space-time with the metric $g_{\mu \nu}^{\prime}=e^{2(\Omega +\omega )}\eta_{\mu \nu}$ is expanded in terms of the Fock space basis of the space-time with the metric $g_{\mu \nu}=e^{2\Omega}\eta_{\mu \nu}$ there appears the state of the sum of the Dirac sea and a state called $|\text{pair}\rangle_{[\tilde{\Psi},\Omega]}$. In particular, in the case of $\Omega =0$, the state $|\text{pair}\rangle_{[\tilde{\Psi},\Omega =0]}$ describes the one of the particle-antiparticle pairs. In perturbation theory, between the formulation in terms of the empty vacuum and the formulation in terms of the Dirac sea in which the negative energy states are all filled, there is no difference in the result. However in the case of the nonperturbative phenomena such as the particle-antiparticle pair production, there appears apparent difference as we have shown. The origin of this comes from the fact that the vacuum is not empty but there is a sea of the infinitely many negative energy particles and these are transformed under the Weyl transformation.

When the Dirac sea, which is defined in a diffeomorphism invariant manner, is acted by the Weyl transformation, the pairs of the particle and antiparticle are created and the Weyl symmetry is broken. We have shown that this is precisely the origin of the Weyl anomaly. In the Fujikawa's method based on the path integral formalism~\cite{rf:fujikawa}, the path integral measure violates the symmetry and the Weyl anomaly stems from the Jacobian of the Weyl transformation. In our operator formalism, the pair production from the Dirac sea on each time-slice, which occurs during the transition from $-\infty$-past to $+\infty$-future, breaks the Weyl symmetry. The Weyl anomaly is obtained by summing up the breaking effects over all the time.

In this manner, we show the formulation of the physical relation between the pair production state and the appearance of the Weyl anomaly. Our formulation can be applied to many physical phenomena, irrelevant to the detail of the models. For instance the state expression of the pair production can be used by changing the gravitational field to the electromagnetic field to investigate the dynamics of the electrons and holes in the materials. As for the physics of the high energy particle and the universe we are under investigation and some of the topics will be shown in the following section \ref{sec3.2}.

\subsection{Outlooks}\label{sec3.2}

Really the main motivation for the present work is the hope that we may extend our formulation to slightly different but related issues:
\begin{enumerate}
\item[(1)] The pair production of bosons to generate the Weyl anomaly: 

In our previous works \cite{rf:hole-th}, we formulated the boson sea which is the vacuum being filled with negative energy bosons. If we set the background metric to the conformally flat one and transform the boson sea by the Weyl transformation, we can construct the pair produced state from the boson sea vacuum, which would be the bosonic version of (\ref{eq:2-weyl.dirac.a}). All we should do may be parallel to the method shown in the present article.
\item[(2)] All other anomalies such as the gravitational anomaly being generated via the pair production from the sea: 

We can apply our method to the physical interpretations of all anomalies and unify them in terms of the pair production from the sea vacuum.
\item[(3)] The Hawking radiation and the gravitational wave: 

When we consider the sea vacuum in the Schwarzschild or de Sitter space-time background, we can express the Hawking radiation as the pair production state. In particular, in the case of the complex boson and the boson sea, we can represent the gravitational wave by using the pair production state as the Hawking radiation, because the number of the degrees of freedom of the gravitational wave is two which is identical to that of the complex boson. Then we may be able to evaluate the entropies of the black hole and the inflationary universe using the notion of the typicality \cite{rf:typicality} or the formulation of statistical mechanics based on the thermal pure quantum states \cite{rf:tpq}.
\end{enumerate}

\section*{Acknowledgement}
The authors acknowledge Yukinori Nagatani for useful discussion in the early stage of the present work. One of us (H.B.N) thanks for hospitality to stay at the Niels Bohr Institute as emeritus professor. One of the authors (M.N) would like to thank the Niels Bohr Institute for hospitality extended to him during his stay at N.B.I. M.N is supported by the JSPS Grant in Aid for Scientific Research No. 24540293.

\appendix

\section{Invariance of the Dirac sea under the diffeomorphism}\label{app1}

The rescaling (\ref{eq:2-rescale}) in Section \ref{sec2} is meant that Dirac sea becomes invariant under diffeomorphism. In this appendix we give a proof of the invariance. 

Under the infinitesimal coordinate transformation 
\begin{align}
	x^{\prime \mu}=x^{\mu}-\epsilon^{\mu}(\vec{x})
	\label{eq:app-coordinate.trans}
\end{align}
the rescaled fermion field $\tilde{\Psi}=\sqrt[4]{g}\Psi$ transforms as 
\begin{align}
	\tilde{\Psi}^{\prime}(\vec{x})
	=\tilde{\Psi}(\vec{x})
	+\epsilon^{\mu}\partial_{\mu}\tilde{\Psi}(\vec{x})
	+\frac{1}{2}(\partial_{\mu}\epsilon^{\mu})\tilde{\Psi}(\vec{x}). 
	\label{eq:app-fermion.trans}
\end{align}
As is done for (\ref{eq:2-weyl.trans-relation}) the transformation rules of the creation and annihilation operators under the diffeomorphism (\ref{eq:app-fermion.trans}) is written as 
\begin{align}
	b_n^{[\tilde{\Psi}^{\prime},\Omega]}
	& =\oint \frac{dx}{2\pi}\> e^{-\Omega (\vec{x})}\cdot 
	e^{\frac{1}{2}\Omega (\vec{x})}e^{n(t+ix)}
	\left(1+\epsilon^{\mu}\partial_{\mu}
	+\frac{1}{2}\partial_{\mu}\epsilon^{\mu}\right)
	\sum_{m=-\infty}^{+\infty}b_m^{[\tilde{\Psi},\Omega]}
	\frac{e^{\frac{1}{2}\Omega (\vec{x})}}{e^{m(t+ix)}} \nonumber \\
	& \equiv \sum_{m=-\infty}^{+\infty}D_{n,m}b_m^{[\tilde{\Psi},\Omega]}, 
	\nonumber \\
	\bar{b}_n^{[\tilde{\Psi}^{\prime},\Omega]}
	& =\oint \frac{dx}{2\pi}\> e^{-\Omega (\vec{x})}\cdot 
	e^{\frac{1}{2}\Omega (\vec{x})}e^{n(t-ix)}
	\left(1+\epsilon^{\mu}\partial_{\mu}
	+\frac{1}{2}\partial_{\mu}\epsilon^{\mu}\right)
	\sum_{m=-\infty}^{+\infty}\bar{b}_m^{[\tilde{\Psi},\Omega]}
	\frac{e^{\frac{1}{2}\Omega (\vec{x})}}{e^{m(t-ix)}} \nonumber \\
	& \equiv \sum_{m=-\infty}^{+\infty}\bar{D}_{n,m}
	\bar{b}_m^{[\tilde{\Psi},\Omega]}. 
	\label{eq:app-diffeo-relation}
\end{align}
Here we defined the representation matrices of the infinitesimal diffeomorphism as 
\begin{align}
	D_{n,m}
	& \equiv \delta_{n,m}+\oint dx\> 
	\langle n|x\rangle_{[\Omega]}
	\left(\epsilon^{\mu}\partial_{\mu}
	+\frac{1}{2}\partial_{\mu}\epsilon^{\mu}\right)
	{}_{[\Omega]}\langle x|m\rangle , \nonumber \\
	\bar{D}_{n,m}
	& \equiv \delta_{n,m}+\oint dx\> 
	\langle \bar{n}|x\rangle_{[\Omega]}
	\left(\epsilon^{\mu}\partial_{\mu}
	+\frac{1}{2}\partial_{\mu}\epsilon^{\mu}\right)
	{}_{[\Omega]}\langle x|\bar{m}\rangle .
	\label{eq:app-diffeo-matrix}
\end{align}
By using these expressions, we can estimate the transformation law of the Dirac sea under the diffeomorphism up to the first order of the infinitesimal parameter $\epsilon$ as 
\begin{align}
	|\text{sea}\rangle_{[\tilde{\Psi}^{\prime},\Omega](t)}^D
	& =\prod_{n\geq 0}b_n^{[\tilde{\Psi}^{\prime},\Omega]\dagger}|0\rangle 
	\otimes 
	\prod_{\bar{n}<0}
	\bar{b}_{\bar{n}}^{[\tilde{\Psi}^{\prime},\Omega]\dagger}
	|\bar{0}\rangle \nonumber \\
	& =\prod_{n\geq 0}\left(\sum_{m=-\infty}^{+\infty}D_{n,m}^{\dagger}
	b_m^{[\tilde{\Psi},\Omega]\dagger}\right)|0\rangle \otimes 
	\prod_{\bar{n}<0}\left(\sum_{\bar{m}=-\infty}^{+\infty}
	\bar{D}_{\bar{n},\bar{m}}^{\dagger}
	\bar{b}_{\bar{m}}^{[\tilde{\Psi},\Omega]\dagger}\right)|\bar{0}\rangle 
	\nonumber \\
	& =\det \left[\delta_{n,m}+\oint dx\> 
	{}_{[\Omega](t)}\langle x|n\rangle 
	\left(\epsilon^{\mu}\partial_{\mu}
	+\frac{1}{2}\partial_{\mu}\epsilon^{\mu}\right)
	\langle m|x\rangle_{[\Omega](t)}\right]_{n,m\geq 0} \nonumber \\
	& \qquad \times \det \left[\delta_{\bar{n},\bar{m}}
	+\oint dx\> 
	\langle \bar{n}|x\rangle_{[\Omega](t)}
	\left(\epsilon^{\mu}\partial_{\mu}
	+\frac{1}{2}\partial_{\mu}\epsilon^{\mu}\right)
	{}_{[\Omega](t)}\langle x|\bar{m}\rangle \right]_{\bar{n},\bar{m}>0}|
	\text{sea}\rangle_{[\tilde{\Psi},\Omega](t)} \nonumber \\
	& \quad +|\widetilde{\text{pair}}\rangle_{[\tilde{\Psi},\Omega](t)}^D 
	\nonumber \\
	& =\exp \Bigg[\sum_{n\geq 0}\oint dx\> 
	\bigg\{{}_{[\Omega](t)}\langle x|n\rangle 
	\left(\epsilon^{\mu}\partial_{\mu}
	+\frac{1}{2}\partial_{\mu}\epsilon^{\mu}\right)
	\langle n|x\rangle_{[\Omega](t)} \nonumber \\
	& \qquad \qquad \qquad \qquad \quad 
	+\langle \bar{n}|x\rangle_{[\Omega](t)}
	\left(\epsilon^{\mu}\partial_{\mu}
	+\frac{1}{2}\partial_{\mu}\epsilon^{\mu}\right)
	{}_{[\Omega](t)}\langle x|\bar{n}\rangle \bigg\}\Bigg]
	|\text{sea}\rangle_{[\tilde{\Psi},\Omega](t)} \nonumber \\
	& \quad +|\widetilde{\text{pair}}\rangle_{[\tilde{\Psi},\Omega](t)}^D. 
	\label{eq:app-diffeo.dirac.a}
\end{align}
Then, when we consider the vacuum-to-vacuum amplitude, we obtain the following factor $J_{\omega}^D$ for the diffeomorphism instead of $J_{\omega}$ in (\ref{eq:2-weyl.dirac.c'}) for the Weyl transformation. We also change the basis to $\left\{\chi_r(\vec{x})\right\}$ and finally 
\begin{align}
	J_{\omega}^D
	& =\exp \left[\sum_{r=0}^{+\infty}\int d^2\vec{x}\> 
	\left\{\chi_r(\vec{x})\left(\epsilon^{\mu}\partial_{\mu}
	+\frac{1}{2}\partial_{\mu}\epsilon^{\mu}\right)
	\chi_r^{\dagger}(\vec{x})
	+\chi_r^{\dagger}(\vec{x})\left(\epsilon^{\mu}\partial_{\mu}
	+\frac{1}{2}\partial_{\mu}\epsilon^{\mu}\right)\chi_r(\vec{x})\right\}
	\right] \nonumber \\
	& =\exp \left[\sum_{r=0}^{+\infty}\int d^2\vec{x}\> \partial_{\mu}
	\left\{\epsilon^{\mu}\chi_r^{\dagger}(\vec{x})\chi_r(\vec{x})\right\}
	\right] \nonumber \\
	& =1. 
	\label{eq:app-diffeo.dirac}
\end{align}
In this way the contribution from the Dirac sea to the potential anomaly like the identity (\ref{eq:2-identity}) vanishes as total derivative. 

\section{Invariance of the Dirac sea under the conformal transformation}\label{app2}

In this appendix B, we consider the transformation of Dirac sea under the conformal transformation. As is explained briefly below, in the conformal field theory, it is a field theory on the flat space-time. Thus it may be reasonable to suppose that the conformal transformation does not contribute to the pair production unlike to the case of the Weyl transformation in curved background space-time. 

Firstly we make the action (\ref{eq:1-diffeo.action}) by diffeomorphism such that the metric tensor becomes the form of the conformal flat (\ref{eq:1-conformal.metric}). We then use the complex coordinate $(z,\bar{z})$ by making the coordinate transformation, 
\begin{align}
	z=e^w=e^{x^0+ix^1},\quad \bar{z}=e^{\bar{w}}=e^{x^0-ix^1}. 
	\label{eq:app-complex.coordinate}
\end{align}
The metric (\ref{eq:1-conformal.metric}) is assumed to be of the form of the product of holomorphic and anti-holomorphic part, 
\begin{align}
	g_{\mu \nu}=e^{\partial \omega (z)
	+\bar{\partial}\bar{\omega}(\bar{z})}\eta_{\mu \nu}. 
	\label{eq:app-conformal.metric}
\end{align}
It is written in terms of the zweibein $e_{i\mu}(\vec{x})$ such that 
\begin{align}
	e_{zz}(z)=\frac{1}{2}e^{\partial \omega (z)},\quad 
	e_{\bar{z}\bar{z}}(\bar{z})
	=\frac{1}{2}e^{\bar{\partial}\bar{\omega}(\bar{z})},\quad 
	e_{z\bar{z}}=e_{\bar{z}z}=0. 
	\label{eq:app-zweibein}
\end{align}
Then the action reads 
\begin{align}
	S & =\frac{1}{4\pi}\int dzd\bar{z}\> 
	\left[
	\left\{e^{\frac{1}{2}\partial \omega (z)}\psi^{\dagger}\right\}
	\bar{\partial}
	\left\{e^{\frac{1}{2}\partial \omega (z)}\psi \right\}
	+\left\{e^{\frac{1}{2}\bar{\partial}\bar{\omega}(\bar{z})}
	\bar{\psi}^{\dagger}\right\}
	\partial 
	\left\{e^{\frac{1}{2}\bar{\partial}\bar{\omega}(\bar{z})}
	\bar{\psi}\right\}
	\right]. 
	\label{eq:app-action}
\end{align}
The action (\ref{eq:app-action}) is further reduced to be of the flat space-time one 
\begin{align}
	S & =\frac{1}{4\pi}\int dzd\bar{z}\> 
	\left(\psi^{\dagger}\bar{\partial}\psi 
	+\bar{\psi}^{\dagger}\partial \bar{\psi}\right)
	\label{eq:app-conformal.action}
\end{align}
by applying the Weyl invariance 
\begin{align}
	\left\{ \begin{array}{l}
	\omega (z)\to \omega^{\prime}(z)=\omega (z)+\xi (z) \\
	\bar{\omega}(\bar{z})\to \bar{\omega}^{\prime}(\bar{z})
	=\bar{\omega}(\bar{z})+\bar{\xi}(\bar{z}) \\
	\psi (z,\bar{z})\to \psi^{\prime}(z,\bar{z})
	=e^{-\frac{1}{2}\partial \xi (z)}\psi (z,\bar{z}) \\
	\bar{\psi}(z,\bar{z})\to \bar{\psi}^{\prime}(z,\bar{z})
	=e^{-\frac{1}{2}\bar{\partial}\bar{\xi}(\bar{z})}\bar{\psi}
	(z,\bar{z}) \\
	\end{array} \right.. 
	\label{eq:app-weyl.invariance}
\end{align}

We now construct the Dirac sea by making the standard canonical second quantization. The field expansion is performed under the periodic boundary condition 
\begin{align*}
	& \psi (t,x+2\pi )=\psi (t,x),\\
	& \bar{\psi}(t,x+2\pi )=\bar{\psi}(t,x)
\end{align*}
and given as 
\begin{align}
	& \psi (z)=\sum_{n=-\infty}^{+\infty}\frac{b_n}{z^{n+\frac{1}{2}}}, 
	\quad 
	\psi^{\dagger}(z)=\sum_{n=-\infty}^{+\infty}b_n^{\dagger}
	z^{n-\frac{1}{2}}, \nonumber \\
	& \bar{\psi}(\bar{z})=\sum_{n=-\infty}^{+\infty}
	\frac{b_n}{\bar{z}^{n+\frac{1}{2}}}, \quad 
	\bar{\psi}^{\dagger}(\bar{z})=\sum_{n=-\infty}^{+\infty}
	b_n^{\dagger}\bar{z}^{n-\frac{1}{2}}. 
	\label{eq:app-solution}
\end{align}
The commutation relations of the creation and annihilation operators are of the usual one 
\begin{align}
	\left\{b_n,b_m^{\dagger}\right\}
	=\left\{\bar{b}_n,\bar{b}_m^{\dagger}\right\}
	=\delta_{n,m}, \quad \text{others}=0. 
	\label{eq:app-commutator}
\end{align}
Thus the Dirac sea is constructed as 
\begin{align}
	& |\text{sea}\rangle =
	\prod_{n\geq 0}b_n^{\dagger}|0\rangle \otimes 
	\prod_{m<0}\bar{b}_m^{\dagger}|\bar{0}\rangle , \nonumber \\
	& \text{s.t. }\> b_n|0\rangle =\bar{b}_n|\bar{0}\rangle =0 
	\quad \text{for all $n$.}
	\label{eq:app-dirac.sea}
\end{align}

Now the action (\ref{eq:app-conformal.action}) is invariant under the following infinitesimal conformal transformation 
\begin{align}
	& \left\{ \begin{array}{l}
	z\to z^{\prime}=z+\epsilon (z) \\
	\bar{z}\to \bar{z}^{\prime}=\bar{z}+\bar{\epsilon}(\bar{z}) \\
	\psi (z)\to \psi^{\prime}(z^{\prime})
	=\left\{1-\frac{1}{2}\partial \epsilon (z)\right\}\psi (z) \\
	\bar{\psi}(\bar{z})\to 
	\bar{\psi}^{\prime}(\bar{z}^{\prime})
	=\left\{1-\frac{1}{2}\bar{\partial}\bar{\epsilon}(\bar{z})\right\}
	\bar{\psi}(\bar{z})
	\end{array} \right..
	\label{eq:app-conformal.invariance}
\end{align}
This transformation can be rewritten as the expansion coefficients of the fields $\psi$ and $\bar{\psi}$. In the same manner as the case of the Weyl transformation (\ref{eq:2-weyled.field.expansion}) and (\ref{eq:2-field.expansion}), the fields before and after the conformal transformation are Laurent expanded by using the same basis as 
\begin{align}
	\psi^{\prime}(z)
	& =\left\{1+\partial \epsilon (z)\right\}^{-\frac{1}{2}}\psi (z)
	-\epsilon (z)\partial \psi (z) \nonumber \\
	& =\left\{1-\frac{1}{2}\partial \epsilon (z)
	-\epsilon (z)\partial \right\}\sum_{n=-\infty}^{+\infty}
	\frac{b_n}{z^{n+\frac{1}{2}}} \nonumber \\
	& \equiv \sum_{n=-\infty}^{+\infty}
	\frac{b_n^{\prime}}{z^{n+\frac{1}{2}}}, \nonumber \\
	\bar{\psi}^{\prime}(\bar{z})
	& =\left\{1+\bar{\partial}\bar{\epsilon}(\bar{z})
	\right\}^{-\frac{1}{2}}\bar{\psi}(\bar{z})-\bar{\epsilon}(\bar{z})
	\bar{\partial}\bar{\psi}(\bar{z}) \nonumber \\
	& =\left\{1-\frac{1}{2}\bar{\partial}\bar{\epsilon}(\bar{z})
	-\bar{\epsilon}(\bar{z})\bar{\partial}\right\}
	\sum_{n=-\infty}^{+\infty}\frac{\bar{b}_n}{\bar{z}^{n+\frac{1}{2}}} 
	\nonumber \\
	& \equiv \sum_{n=-\infty}^{+\infty}
	\frac{\bar{b}_n^{\prime}}{\bar{z}^{n+\frac{1}{2}}}.
	\label{eq:app-laurent.conformal}
\end{align}
We then obtain the transformation rule of $b_n$ and $\bar{b}_n$, 
\begin{align}
	b_n^{\prime}
	& =\oint \frac{dz}{2\pi i}\> z^{n-\frac{1}{2}}
	\left\{1-\frac{1}{2}\partial \epsilon (z)
	-\epsilon (z)\partial \right\}\sum_{m=-\infty}^{+\infty}
	\frac{b_m}{z^{m+\frac{1}{2}}} \nonumber \\
	& \equiv \sum_{m=-\infty}^{+\infty}C_{n,m}b_m, \nonumber \\
	\bar{b}_n^{\prime}
	& =-\oint \frac{d\bar{z}}{2\pi i}\> 
	\bar{z}^{n-\frac{1}{2}}
	\left\{1-\frac{1}{2}\bar{\partial}\bar{\epsilon}(\bar{z})
	-\bar{\epsilon}(\bar{z})\bar{\partial}\right\}
	\sum_{m=-\infty}^{+\infty}\frac{\bar{b}_m}{\bar{z}^{m+\frac{1}{2}}} 
	\nonumber \\
	& \equiv \sum_{m=-\infty}^{+\infty}\bar{C}_{n,m}\bar{b}_m. 
	\label{eq:app-conformal.trans-relation}
\end{align}
Here the matrix representation of the conformal transformation is defined by 
\begin{align}
	C_{n,m} 
	& \equiv \delta_{n,m}-\oint \frac{dz}{2\pi i}\> 
	z^{n-\frac{1}{2}}\left\{\frac{1}{2}\partial \epsilon (z)+\epsilon (z)
	\partial \right\}z^{-m-\frac{1}{2}}, \nonumber \\
	\bar{C}_{n,m} 
	& \equiv \delta_{n,m}+\oint \frac{d\bar{z}}{2\pi i}\> 
	\bar{z}^{n-\frac{1}{2}}\left\{\frac{1}{2}\bar{\partial}
	\bar{\epsilon}(\bar{z})+\bar{\epsilon}(\bar{z})\bar{\partial}\right\}
	\bar{z}^{-m-\frac{1}{2}}.
	\label{eq:app-conformal.trans-matrix}
\end{align}
As a byproduct, it is easily shown that (\ref{eq:app-conformal.trans-relation}) is indeed the Bogolyubov transformation in the following: 
\begin{align}
	\left\{b_n^{\prime},\> b_m^{\prime \dagger}\right\}
	& =\left\{\sum_kC_{n,k}b_k,\sum_lC_{m,l}^{\dagger}b_l^{\dagger}\right\}
	=\sum_{k=-\infty}^{+\infty}C_{n,k}C_{m,k}^{\dagger} \nonumber \\
	& =\sum_{k=-\infty}^{+\infty}
	\left[\delta_{n,k}-\oint \frac{dz}{2\pi i}\> 
	z^{n-\frac{1}{2}}\left\{\frac{1}{2}\partial \epsilon (z)+\epsilon (z)
	\partial \right\}z^{-k-\frac{1}{2}}\right] \nonumber \\
	& \qquad \qquad \times 
	\left[\delta_{m,k}-\oint \frac{dz}{2\pi i}\> 
	z^{-m-\frac{1}{2}}\left\{\frac{1}{2}\partial \epsilon (z)+\epsilon (z)
	\partial \right\}z^{k-\frac{1}{2}}\right] \nonumber \\
	& =\delta_{n,m}-\oint \frac{dz}{2\pi i}\> 
	z^{n-\frac{1}{2}}\left\{\frac{1}{2}\partial \epsilon (z)+\epsilon (z)
	\partial \right\}z^{-m-\frac{1}{2}} \nonumber \\
	& \qquad \qquad -\oint \frac{dz}{2\pi i}\> 
	z^{-m-\frac{1}{2}}\left\{\frac{1}{2}\partial \epsilon (z)+\epsilon (z)
	\partial \right\}z^{n-\frac{1}{2}} \nonumber \\
	& =\delta_{n,m}-\oint \frac{dz}{2\pi i}\> 
	\partial \left\{\epsilon (z)z^{n-m-1}\right\} \nonumber \\
	& =\delta_{n,m}. 
	\label{eq:app-commutator.conformal}
\end{align}
Similarly $\bar{b}_n$ satisfies the same property as (\ref{eq:app-commutator.conformal}).

We now consider the transformation of the Dirac sea under conformal transformation. The procedure of the calculation is much the same as that of Weyl transformation. In the vacuum-to-vacuum amplitude pair production state does not contribute as is shown in Weyl transformation so that we abbreviate it in the following: 
\begin{align}
	|\text{sea}^{\prime}\rangle^C
	& =\prod_{n\geq 0}\left(\sum_{m=-\infty}^{+\infty}C_{n,m}^{\dagger}
	b_m^{\dagger}\right)|0\rangle \otimes 
	\prod_{n<0}\left(\sum_{m=-\infty}^{+\infty}\bar{C}_{n,m}^{\dagger}
	\bar{b}_m^{\dagger}\right)|\bar{0}\rangle \nonumber \\
	& =\det \left[\delta_{n,m}-\oint \frac{dz}{2\pi i}\> 
	z^{n-\frac{1}{2}}\left\{\frac{1}{2}\partial \epsilon (z)
	+\epsilon (z)\partial \right\}z^{-m-\frac{1}{2}}\right]_{n,m\geq 0} 
	\nonumber \\
	& \qquad \times 
	\det \left[\delta_{n,m}+\oint \frac{d\bar{z}}{2\pi i}\> 
	\bar{z}^{-n-\frac{1}{2}}
	\left\{\frac{1}{2}\bar{\partial}\bar{\epsilon}(\bar{z})+\bar{\epsilon}
	(\bar{z})\bar{\partial}\right\}\bar{z}^{m-\frac{1}{2}}\right]_{n,m>0}|
	\text{sea}\rangle \nonumber \\
	& \quad +|\text{pair}\rangle^C \nonumber \\
	& =\exp \left[-\sum_{n\geq 0}\oint \frac{dz}{2\pi i}\> 
	z^{n-\frac{1}{2}}\left\{\frac{1}{2}\partial \epsilon (z)
	+\epsilon (z)\partial \right\}z^{-n-\frac{1}{2}}\right] \nonumber \\
	& \qquad \times \exp \left[+\sum_{n>0}\oint \frac{d\bar{z}}{2\pi i}\> 
	\bar{z}^{-n-\frac{1}{2}}
	\left\{\frac{1}{2}\bar{\partial}\bar{\epsilon}(\bar{z})+\bar{\epsilon}
	(\bar{z})\bar{\partial}\right\}\bar{z}^{n-\frac{1}{2}}\right]
	|\text{sea}\rangle +|\text{pair}\rangle^C \nonumber \\
	& =\exp \Bigg[-\sum_{n\geq 0}\oint \frac{dz}{2\pi i}\> 
	\left\{\frac{1}{2}\partial \frac{\epsilon (z)}{z}
	-n\frac{\epsilon (z)}{z^2}\right\} \nonumber \\
	& \qquad \qquad \qquad +\sum_{n>0}\oint \frac{d\bar{z}}{2\pi i}\> 
	\left\{\frac{1}{2}\bar{\partial}
	\frac{\bar{\epsilon}(\bar{z})}{\bar{z}}
	+n\frac{\bar{\epsilon}(\bar{z})}{\bar{z}^2}\right\}\Bigg]
	|\text{sea}\rangle +|\text{pair}\rangle^C. 
	\tag{\ref{eq:app-conformal.dirac}a}
	\label{eq:app-conformal.dirac.a}
\end{align}
In (\ref{eq:app-conformal.dirac.a}) we drop the total derivative term with respect to $z$ and $\bar{z}$ and then insert the Laurent expansion of $\epsilon$ and $\bar{\epsilon}$ such that 
\begin{align*}
	& \epsilon (z)=\sum_n \epsilon_n/z^{n-1}, \\
	& \bar{\epsilon}(\bar{z})=\sum_n \bar{\epsilon}_n/\bar{z}^{n-1}. 
\end{align*}
The equation (\ref{eq:app-conformal.dirac.a}) reads 
\begin{align}
	|\text{sea}^{\prime}\rangle^C
	& =\exp \left[(\epsilon_0-\bar{\epsilon_0})\sum_{n\geq 0}n\right]
	|\text{sea}\rangle +|\text{pair}\rangle^C \nonumber \\
	& =\exp \left[-\frac{i}{12}\text{Im} \epsilon_0\right]
	|\text{sea}\rangle +|\text{pair}\rangle^C. 
	\label{eq:app-conformal.dirac}
\end{align}
In the last equality we used the $\zeta$-function regularization $\sum_{n\geq 0}n=-\frac{1}{12}$. In (\ref{eq:app-conformal.dirac}) the factor in front of the Dirac sea $|\text{sea}\rangle_t$ is nothing but the phase and we may conclude that the conformal transformation corresponds to the unitary transformation except for the pair production state. Furthermore in the factor $\epsilon_0$ and $\bar{\epsilon}_0$ are the zero modes of $\partial \epsilon (z)$ and $\bar{\partial}\bar{\epsilon}(\bar{z})$ respectively. Among them the physical meaningful terms are only the sum $\partial \epsilon (z)+\bar{\partial}\bar{\epsilon}(\bar{z})$ which is the change of the conformal part of the metric. We may set $\text{Im}\epsilon_0=0$. Thus under the conformal transformation we conclude that as far as we concern the vacuum-to-vacuum amplitude (\ref{eq:2-identity}), the Dirac sea does not produce any contribution like the pair production. This result coincides with the fact that the conformal field theory is just a closed theory in the flat space-time.

\section{About the determinants in front of $|\text{sea}\rangle_{[\tilde{\Psi},\Omega](t)}$ in (\ref{eq:2-weyl.dirac.a})}\label{app3}

We start from the second line of (\ref{eq:2-weyl.dirac.a}) and treat the left and right movers independently. For the left mover, we can assume that the creation oeprators $b_{n\geq 0}^{[\tilde{\Psi},\Omega]\dagger}$ of the Dirac sea (see e.g. (\ref{eq:1-dirac.sea})) is arranged in the ascending order as to the index $n$: 
\begin{align}
	|\text{sea}\rangle_{[\tilde{\Psi},\Omega](t)}
	=b_0^{[\tilde{\Psi},\Omega]\dagger}b_1^{[\tilde{\Psi},\Omega]\dagger}
	b_2^{[\tilde{\Psi},\Omega]\dagger}\cdots |0\rangle 
	\otimes \text{[right mover part]}. 
\end{align}
Then, taking notice of $m\geq 0$ of the summation $\sum_m$, cutting off the indices $n$ and $m$ up to finite $M$ and utilizing the fermionic nature of $b_m^{[\tilde{\Psi},\Omega]\dagger}$, we get 
\begin{align}
	\prod_{n=0}^M & \left(\sum_{m=0}^M
	W_{n,m}^{\dagger}b_m^{[\tilde{\Psi},\Omega]\dagger}\right)|0\rangle 
	\nonumber \\
	& =\sum_{\sigma =\text{\tiny $\left( \begin{array}{cccc}
	0 & 1 & \cdots & M \\ i_0 & i_1 & \cdots & i_M 
	\end{array}\right)$}}
	W_{0,i_0}^{\dagger}b_{i_0}^{[\tilde{\Psi},\Omega]\dagger}\times 
	W_{1,i_1}^{\dagger}b_{i_1}^{[\tilde{\Psi},\Omega]\dagger}\times 
	\cdots \times 
	W_{M,i_M}^{\dagger}b_{i_M}^{[\tilde{\Psi},\Omega]\dagger}|0\rangle 
	\nonumber \\
	& =\left[\sum_{\sigma =\text{\tiny $\left( \begin{array}{cccc}
	0 & 1 & \cdots & M \\ i_0 & i_1 & \cdots & i_M 
	\end{array}\right)$}}\text{sign} (\sigma )\> 
	W_{0,i_0}^{\dagger}W_{1,i_1}^{\dagger}\cdots W_{M,i_M}^{\dagger}\right]
	b_0^{[\tilde{\Psi},\Omega]\dagger}b_1^{[\tilde{\Psi},\Omega]\dagger}
	\cdots b_M^{[\tilde{\Psi},\Omega]\dagger}|0\rangle \nonumber \\
	& =\left(\det W_{n,m}^{\dagger}\right)_{n,m\geq 0}
	b_0^{[\tilde{\Psi},\Omega]\dagger}b_1^{[\tilde{\Psi},\Omega]\dagger}
	\cdots b_M^{[\tilde{\Psi},\Omega]\dagger}|0\rangle . 
\end{align}
The terms about $m<0$ become the states in the fifth line of (\ref{eq:2-weyl.dirac.a}). The same procedure can be followed for the righr mover. After taking the limit $M\to +\infty$, we obtain (\ref{eq:2-weyl.dirac.a}). The determinants in (\ref{eq:2-identity_e}) can be obtain in the same way except for the continuum index $x$.



\end{document}